%% file: Drone_Comm.tex
\documentclass[journal,comsoc]{IEEEtran}

\usepackage[font=small]{caption}
\usepackage[labelsep=period]{caption}
\usepackage{float}
\usepackage[latin1]{inputenc}
\usepackage{graphicx}
\usepackage[cmex10]{amsmath}
\usepackage[margin=2cm,nohead]{geometry}
\usepackage[nospace,noadjust]{cite}
\usepackage{amsfonts}
\usepackage{amsmath}
\usepackage{amsthm}
\usepackage{mathrsfs}
\usepackage{amssymb,mathrsfs}
\usepackage[mathscr]{euscript}
\usepackage{subcaption}
\usepackage{epstopdf} 
\usepackage[dvipsnames]{xcolor}
\usepackage{pgfplots}
\pgfplotsset{compat=newest}
\usetikzlibrary{plotmarks}
\usepackage{grffile}
\usepackage{amsmath}
\usepackage[normalem]{ulem}
\usepackage{gensymb}
\usepackage{balance}

\usepackage{balance}

\def\plos{\mathcal{P}_\text{LoS}}

\pgfplotsset{every axis/.append style={
        scaled x ticks = false, 
        x tick label style={/pgf/number format/.cd, fixed, fixed zerofill,
                            int detect, 1000 sep={},precision=3}
    }
}

\begin{document}
\title{Key Technologies and System Trade-Offs for Detection and Localization of Amateur Drones}

\author{Mohammad Mahdi Azari, Hazem Sallouha, Alessandro Chiumento, \\  Sreeraj Rajendran, Evgenii Vinogradov, and Sofie Pollin \\ \vspace{3mm}  Email: mahdi.azari@kuleuven.be
} 
\maketitle

\begin{abstract}
The use of amateur drones (ADrs) is expected to significantly increase over the upcoming years. However, regulations do not allow such drones to fly over all areas, in addition to typical altitude limitations. As a result, there is an urgent need for ADrs surveillance solutions. These solutions should include means of accurate detection, classification, and localization of the unwanted drones in a \textit{no-fly zone}. In this paper, we give an overview of promising techniques for modulation classification and signal strength based localization of ADrs by using surveillance drones (SDrs). By introducing a generic altitude dependent propagation model, we show how detection and localization performance depend on the altitude of SDrs. Particularly, our simulation results show a 25\,dB reduction in the minimum detectable power or 10 times coverage enhancement of an SDr by flying at the optimum altitude. Moreover, for a target \textit{no-fly zone}, the location estimation error of an ADr can be remarkably reduced by optimizing the positions of the SDrs. Finally, we conclude the paper with a general discussion about the future work and possible challenges of the aerial surveillance systems.
\end{abstract}

\section{Introduction}
Detecting the presence of amateur drones (ADrs) in a \textit{no-fly zone} is an arduous task as the ADrs are usually small objects flying at low altitudes. These ADrs can pose great safety and security problems in critical locations such as power plants, military zones, densely populated areas and private residences \cite{Solodov2017_ThreatNuclear}. Such a variety of \textit{no-fly zone} locations requires a flexible surveillance solution. The research community is very active in developing techniques to determine the location and subsequently track unidentified drones \cite{Rozantsev_GroundCamera,Kim_sound_2017, Hoffmann_2016_Radar,Nguyen:RFDETECTION}.

Current solutions can be divided into active methods, in which a ground infrastructure actively scans the \textit{no-fly zone} for intruders using video or radar techniques \cite{Hoffmann_2016_Radar,Rozantsev_GroundCamera}, or passive methods in which an ADr is detected by its RF transmission or audio signature \cite{Nguyen:RFDETECTION,Kim_sound_2017}. 
Active Radar-based methods require large mono- or multi- static RF nodes used to scan the \textit{no-fly zone}. Frequency sweeping is used to scan for the presence of drones and classification techniques are employed to determine the nature of the detected drone and to track it. Such solutions are very powerful but require the purchase of large devices with fixed coverage radius and the small size of the ADrs poses great detection limitations \cite{Hoffmann_2016_Radar}. Solutions based on video detection require the presence of either distributed camera equipment or $360\degree$ video recording devices \cite{Rozantsev_GroundCamera}. Video processing is then applied on the recorded image to identify whether a drone has entered the protected space, the drone is then identified and tracked.

Passive methods, on the other hand, listen to either the ADr emitted audio or to the transmitted RF signal (i.e. control or downlink to the ADr's base station). Sound-based solutions are able to identify the presence and model of commercial ADrs by listening for specific motor sound signatures but require distributed microphone arrays to track the drone \cite{Kim_sound_2017}. Passive RF solutions listen instead for the ADrs downlink transmission of, for example, a video stream and are able to localize the drone by determining the source point of the transmission \cite{Nguyen:RFDETECTION}. The main advantages and disadvantages of such methods are listed in Table \ref{tb:Drone_detection}.

\begin{table*}[t]
\centering
\caption{Drone Detection Technologies}
\begin{tabular}{|c|l|l|}
\hline
\multicolumn{1}{|c|}{\textbf{Technology}} & \multicolumn{1}{c|}{\textbf{Advantages}}  & \multicolumn{1}{c|}{\textbf{Disadvantages}} \\ \hline \hline
Video \cite{Rozantsev_GroundCamera} & $\begin{array}{l}
 \textrm{ Mature technology } \\ 
 \textrm{ Accessible equipment}
\end{array}$ & $\begin{array}{l}
 \textrm{ Distributed high resolution camera network or } \\ 
 \textrm{ $3$D cameras needed }\\
 \textrm{ Subject to poor visibility problems }\\
 \textrm{ Fixed installation}
\end{array}$  \\ \hline
Audio \cite{Kim_sound_2017} & $\begin{array}{l}
 \textrm{ Cheap sensors } \\ 
 \textrm{ Limited processing necessary }\\ 
 \textrm{ Accessible equipment}
\end{array}$ & $\begin{array}{l}
 \textrm{ Sensitive to ambient noise}\\
 \textrm{ Limited range}\\ 
 \textrm{ Distributed microphone array necessary}\\
 \textrm{ Large datasets necessary for training}\\
 \textrm{ Fixed installation}
\end{array}$ \\ \hline
Radar \cite{Hoffmann_2016_Radar} & $\begin{array}{l}
 \textrm{ Easily installable} \\ 
\end{array}$ & $\begin{array}{l}
 \textrm{ Either mono-static with limited resolution or} \\ 
 \textrm{ Multi-static and distributed}\\
 \textrm{ Small drone cross-section can be challenging to detect}\\
 \textrm{ Expensive}
\end{array}$ \\ \hline
RF \cite{Nguyen:RFDETECTION} & $\begin{array}{l}
 \textrm{ Easily installable} \\ 
 \textrm{ Cheap sensors}
\end{array}$ & $\begin{array}{l}
 \textrm{ Requires good SNR to perform detection} \\ 
 \textrm{ Susceptible to interference}\\
 \textrm{ Range is limited by link quality}
\end{array}$ \\ \hline
\end{tabular}
\label{tb:Drone_detection}
\end{table*}

The ADrs detection and tracking solutions described above assume the presence of a ground infrastructure. Hence, in densely built-up areas, the number of deployed sensor nodes must be dramatically increased to maintain the required sensor system performance in challenging propagation environments. However, an alternative solution is to detect and localize the ADrs by using surveillance drones (SDrs) with passive RF sensing ability. Using SDrs flying at higher altitudes than the ADrs allows for a flexible solution able to be deployed quickly, it can be used to cover a wide range of ground
surfaces and subsequently, is able to detect and localize the ADrs with high accuracy due to better
propagation conditions at high altitudes (i.e. higher signal-to-noise ratio and line-of-sight probability).

Received signal strength (RSS) localization has been used successfully for the positioning of ground nodes \cite{Hazem}. A good link between the receiving sensor and the node to be localized is necessary to allow RSS-based methods. In fact, it has been shown that the presence of an aerial based sensor, together with an appropriate channel model, can greatly benefit the connection to and thus the localization of target nodes \cite{azari2016joint,azari2017ultra}. The usage of aerial RF monitoring devices can then make use of better channels between ADrs flying at low altitudes and surveillance drones (SDrs) for detection and localization to improve current passive RF solutions. 

\begin{figure*}[t]
	\centering	\includegraphics[width=0.8\textwidth]{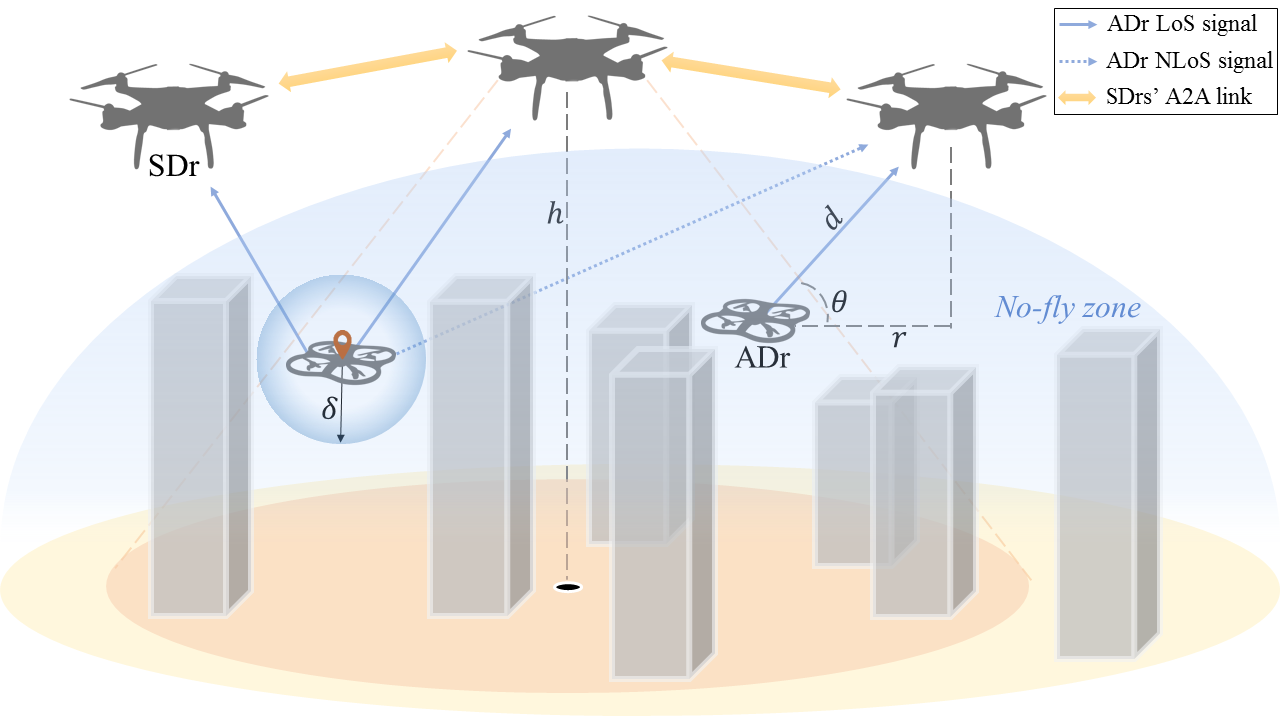} 
	\caption{Due to the high LoS probability $\plos$ at high altitudes, the use of SDrs guarantees higher SNR for better detection and less shadowing for accurate localization compared to terrestrial surveillance. As the ADrs' transmit power is unknown the estimated location of the ADr is represented by a sphere of radius $\delta$.}
	\label{loca}
\end{figure*}

In this paper, we present a framework for the detection and localization of ADrs using SDrs. The proposed framework is based on the received signal by at least three SDrs, see Figure \ref{loca}. As the SDrs are assumed to be flying at higher altitudes than the ADrs, this gives them a better view of the \textit{no-fly zone} and places them much further away from buildings and people. Moreover, they could also be tethered to guarantee increased safety and operated by certified pilots or authorities. The intercepted drone needs to be first identified as such, then by employing a realistic propagation model, the approximate location of the ADr is obtained by multilateration based on the RSS at the SDrs.

The main contributions of this work can be summarized as follows:
\begin{itemize}
\item An overview of the passive RF scanning for detection solutions, relying on recent innovations in deep learning, as a possible solution for the detection and localization of an ADr entering the \textit{no-fly zone},
\item a characterization of the propagation channel seen by an SDr in urban environments, necessary to determine the density of the surveillance sensors for meeting detection and localization constraints,
\item an investigation on the impact of SDr altitude on the coverage area for ADr detection and the range of its detectable transmit power, 
\item a study of the optimal SDrs positioning for better sensing and higher localization accuracy over a target zone based on a range of possible ADr's transmit power,
\item and an overview of the future research directions and challenges.
\end{itemize} 

\section{Passive RF sensing, Technology detection and Localization} \label{passive} 
In this section, an overview of passive RF sensing and localization techniques is presented. For each task, we illustrate a promising solution: deep learning for detection and RSS based distance estimation for localization. Analyses show that detection and localization performance strongly depends on the received signal strength. 

\subsection{Passive RF sensing and detection}

A first step in the surveillance of ADrs is detecting their presence in the \textit{no-fly zone.} When detection of small ADrs with low transmit powers is of interest, it is needed to rely on a very dense infrastructure of RF sensors that are constantly monitoring the radio spectrum. To facilitate detection, we foresee passive RF sensing as a simple yet robust monitoring technique to protect the target \textit{no-fly zone}. Continuous passive radio spectrum monitoring should be enabled in SDrs to allow effective detection of ADrs. For simplifying the analysis, ADrs are assumed to have omni-directional antennas enabling a LoS component to the SDrs as shown in Figure~\ref{loca}. Accordingly, the ADr's antenna gain is identical for any elevation angle $\theta$.

A few recent approaches try to tackle the RF spectrum monitoring problem in a crowd-sourced fashion \cite{electrosense}. Wireless spectrum sensing at high altitude is less challenging as more LoS channels are available when compared to sensing at ground stations. Even with a perfect RF spectrum scanning architecture in place, huge effort is involved in analyzing, detecting and locating transmissions or anomalies in the sensed RF spectrum. Automated systems should be in place to detect authorized or unauthorized transmissions. The detected signals should be then classified to understand the type of transmission. Subsequently, this technology classification can further aid signal power estimation and RSS based localization algorithms.

Accurate technology classification can be achieved to some extent using state-of-the-art (SoA) machine learning classifier models \cite{o2016convolutional, baseline}. To validate technology classification, a few deep learning based time domain models employing convolutional neural networks~(CNN) and long short term memory~(LSTM) units are tested. The deep learning models take IQ samples as input giving out the probability of the data belonging to a particular technology class. Analyses show that the proposed models \cite{o2016convolutional, baseline} yields an average classification accuracy close to 90\% for 11 different technologies, at varying signal-to-noise ratio (SNR) conditions ranging from 0\,dB to 20\,dB, independent of channel characteristics. 
It was also noticed that most of the technology classification algorithms including deep learning solutions require an SNR above 0\,dB for accurate classification which is used as a required SNR threshold in our simulation. 

\subsection{Localization challenges}

The multilateration process is the most prominent method for accurately determining the position of a transmitting source. It is basically a process that uses the estimated distances from such source to at least three different receivers in order to perform localization. Time-based or RSS-based techniques are used to estimate the distance between a transmitter and a receiver \cite{Hazem}. Time-based techniques estimate the distance by multiplying the time of flight by the speed of light. However, defining the time of flight is the bottleneck of these techniques as a very accurate time synchronization is required between the transmitter and the receivers. Certainly, such technique is not feasible for the SDrs since the time of flight cannot be accurately defined as there is no cooperation with the ADr. 

In contrast to time-based techniques, RSS-based solutions are known for their computational simplicity and for not requiring any time synchronization. RSS-based techniques estimate the distance by using a deterministic function that represent RSS as a function of distance. In fact, assuming a known transmit power, the path loss model is a well known representation for the RSS-distance relation. Generally, the major issue limiting the accuracy of RSS-based techniques is the presence of shadowing between the transmitter and the receiver which causes either over- or under-estimation of the distance. However, this drawback can be overcome by using SDrs combined with their altitude optimization which is thoroughly discussed in the next sections. Another challenge that SDrs have to tackle is the unknown transmit power of the ADrs (P$_{\text{Tx}}$) due to the use of different standard of communication (LTE, WiFi, ...) or power control in the ADr. The unknown transmit power leads to an estimated location with uncertainty. As shown in Figure \ref{loca} the uncertainty can be modeled as a sphere of radius $\delta$. This uncertainty can be minimized by classifying the technology being used by the ADrs due to the fact that known technologies have a standard range of transmit powers (e.g., wifi limited to 20\,dBm, LTE limited to 24\,dBm in uplink and 43\,dBm--48\,dBm in downlink, etc). A thorough discussion of SDr altitude's effect on the channel characteristics including path loss and shadowing effects is presented in the following section.

\section{Why Fly Higher: a Better LoS Experience} \label{channel}

To analyze the deployment of SDrs over urban environments, as our main focus, and to characterize the received SNR, a comprehensive understanding of the communication links' channel characteristics is needed. Although air-to-air (A2A) links are dominated by Line-of-Sight (LoS) propagation, the impact of multipath fading due to ground/buildings reflections cannot be ignored. In \cite{Goddemeier2015}, the Rician model with an altitude-dependent K-factor was used to model A2A channels. Naturally, the influence of LoS grows with the increasing ADr's altitude as well as the Doppler frequency due to the higher relative velocity. 

In general, a ground-to-air (G2A) link encounters obstructions between the terrestrial and aerial nodes which limits the LoS and lowers the quality of the channel. Therefore, a G2A link represents a worst case scenario which occurs when the ADrs fly in very low altitudes. 

A G2A channel is observed to be significantly different than A2A and ground-to-ground (G2G) communication due to the high impact of altitude on the channel parameters including path loss exponent, small-scale fading and shadowing effect. To clarify this fact, we consider two extreme cases for a given ground terminal that aims to communicate with a drone seen by an elevation angle of $\theta$:
\begin{enumerate}
\item $\theta \rightarrow 0$: In this case, which is equivalent to $h \rightarrow 0$ (for $r \neq 0$), the channel behavior follows G2G models where the presence of many obstacles results in a dramatic drop for the received power. This significant power decay is reflected onto the channel model by proposing a large path loss exponent $\alpha$ and severe shadowing and small-scale fading effects. It is worth noting that for this case the channel between a transmitter and receiver is roughly always non-line-of-sight (NLoS) as the probability of line-of-sight (LoS) $\plos$ 
converges to zero\footnote{Please note that for a short distance between a transmitter and receiver, the LoS probability $\plos$ exponentially decreases as the link length increases. This impact, however, is approximately neglected for long G2A communication links.}. In fact the LoS probability $\plos$ can be obtained as follows \cite{azari2016joint}
\begin{equation}
\plos(\theta) = \frac{1}{1 + a_0 \exp{(-b_0 \theta)}},
\end{equation}
where $a_0$ and $b_0$ are environment dependent constants.

\item $\theta \rightarrow 90^o$: In this case, which is equivalent to $h \rightarrow \infty$ (for $r \neq 0$), the probability of LoS $\plos$ converges to one  and the channel adopts roughly free space characteristics. Accordingly, a lower path loss exponent and a lighter small-scale fading and shadowing effects are experienced since the environment between the transmitter and receiver becomes less obstructed \cite{azari2016joint}.     
\end{enumerate}

The above-mentioned intuition encourages to model the drone communication channel dependent on the elevation angle as this, easily observable variable, presents a strong correlation with the link quality. To this end, the authors in \cite{al2014modeling} studied a statistical propagation model by considering two major groups of received power and their probability of occurrence, namely LoS and dominant non-LoS (NLoS) components. This model captures different urban environment properties and proposes a $\theta$-dependent path loss and shadowing prediction of the communication channel between a terrestrial and an aerial node.

\begin{figure*}[t]
	\centering
	\begin{subfigure}{0.5\textwidth}
		\centering
 		\input{PL_tet_alpha.tex}
		\caption{}
		\label{PLexponent_tet}
	\end{subfigure}%
	\begin{subfigure}{0.5\textwidth}
		\centering
		\input{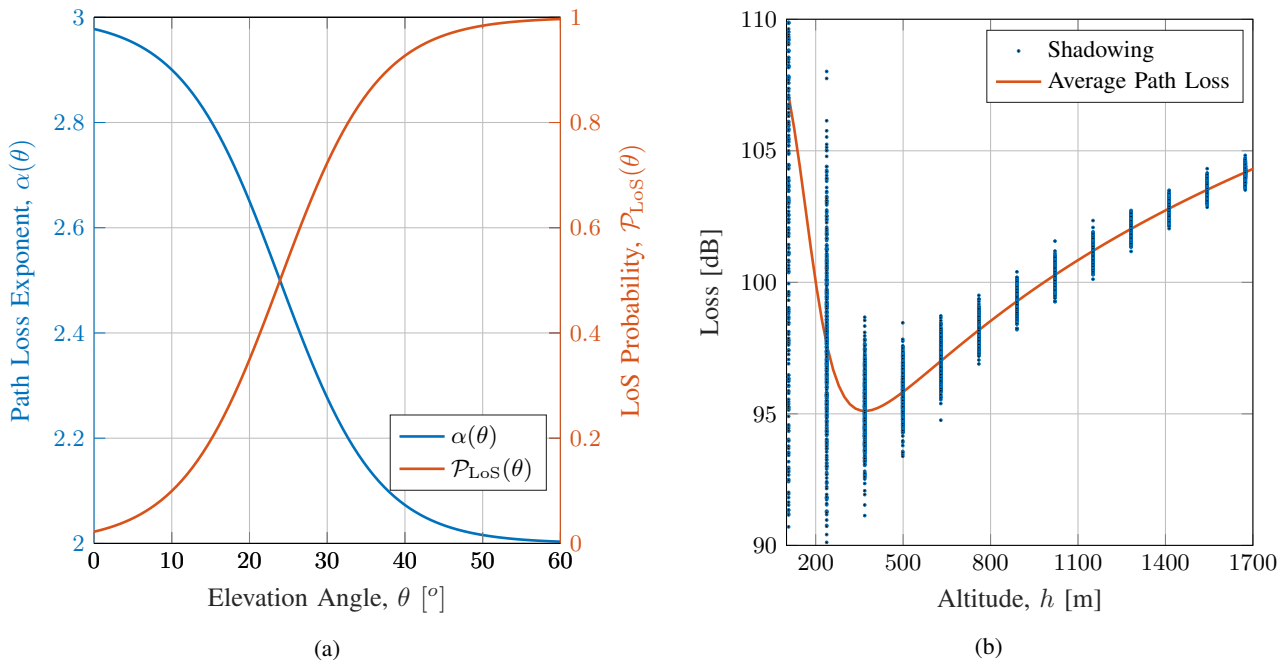}
		\caption{}
		\label{PL_h}
	\end{subfigure}
	\caption{(a) For an urban environment the LoS probability $\plos(\theta)$ is highly dependent on the elevation angle $\theta$. (b) The altitude dependent path loss and shadowing effect for an specific ground node located at $r = 400$\,m.}
	\label{}
 	\vspace{-0.4cm}
\end{figure*}

To extend this G2A model we refer to the work presented in \cite{azari2016joint,azari2017ultra} where we include the small-scale fading and elevation angle dependent path loss exponent. This model unifies a widely used G2G channel model with that of G2A that enables us to study the co-existence of drones with the existing terrestrial networks. In the following, we briefly discuss the dependency of the main components to the elevation angle:
\begin{itemize}
\item \textbf{Path Loss}: path loss exponent is linked to the LoS probability $\plos$ in \cite{azari2016joint} by proposing a negative linear dependency as follows
  \begin{equation}
  \alpha(\theta) = -a_1 \plos(\theta) + b_1,
  \label{plexpo}
  \end{equation}
  where $a_1$ and $b_1$ are environment dependent parameters. Such dependency, illustrated in Figure \ref{PLexponent_tet} for Urban environment, is motivated by the fact that the path loss exponent is proportional to the number of obstacles between a transmitter and receiver. Accordingly, for larger elevation angle the path loss exponent is smaller due to the presence of less obstacles between a ground transmitter and an aerial receiver. The reduction of path loss in Figure \ref{PL_h} is due to a decrease in $\alpha(\theta)$ and the increase is because of an increase in the link length $d$ while the altitude increases.
\item \textbf{Small-Scale Fading}: a G2A link is likely to experience LoS condition and hence Rician fading is an adequate choice for such channel that reflects the combination of LoS and multipath scatters \cite{matolak2017air}. In this model, the fading power is determined by the Rician factor $K$ characterized as the ratio between the power of LoS and multipath components. In fact, the Rician factor represents the severity of fading such that a smaller $K$ corresponds to a more severe fading. Due to a higher LoS probability and the presence of fewer obstacles and scatters at higher altitudes, the average Rician factor could be characterized as a function of $\theta$, i.e. $K = K(\theta)$ \cite{azari2017ultra}. Investigating a functional form for $k(\theta)$ at different urban environment is an open question.
  \item \textbf{Shadowing:} the shadowing effect is studied in \cite{al2014modeling} where a log-normal distribution is considered separately for each LoS and NLoS component. The standard deviation of each group $\sigma_{\text{LoS}}(\theta)$ and $\sigma_{\text{NLoS}}(\theta)$ is characterized using a negative exponential dependency with the elevation angle in which a lower elevation angle and hence altitude leads to a larger variation around the average path loss. Following \cite{al2014modeling}, the overall average shadowing effect in the links can be represented by the standard deviation written as
\begin{equation}
\sigma^2(\theta) = \plos^2(\theta)\cdot\sigma^2_{\text{LoS}}(\theta) + [1-\plos(\theta)]^2\cdot\sigma^2_{\text{NLoS}}(\theta),
\label{shdow}
\end{equation}
which is illustrated in Figure \ref{PL_h}. From the figure, as the drone goes higher the shadowing effect gradually diminishes due to the presence of fewer obstacles between the transmitter and receiver. 
\end{itemize}

By relying on this altitude-dependent shadowing model, it becomes possible to determine the SDr detection coverage and localization accuracy as function of SDr height.  These results include an optimization of SDrs network in order to provide larger coverage, to detect low power ADrs and subsequently maximize the localization accuracy. Finally, please note that the above-mentioned channel characteristics are environment dependent \cite{al2014modeling}, however in the sequel we have examined an Urban environment and focus on the detailed analysis of the performance for this scenario only, rather than quantifying the values for different environments. 

\section{Superior Surveillance Drones} \label{optimization}

We deploy an SDr system for detection and localization, and study its performance as function of altitude. This will give insights into the possibility and benefits of using SDrs for ADrs surveillance. In this section we discuss the efficient positioning of the SDrs to optimally localize an ADr considering the scenario in which the ADr is very close to the ground. In the following, by considering an Urban environment, we link the altitude of the flying SDr to its coverage area, to the ADr's transmit power, and finally to the localization accuracy as the ultimate goal. Note that the detection of ADrs can be done using one SDr whereas localizing them in the \textit{no-fly zone} requires the cooperation of all three SDrs.

\begin{figure*}[t]
	\centering
	\begin{subfigure}{.5\textwidth}
		\centering
		\input{Power_CovRadius.tex}
		\caption{}
		\label{PowGain}
	\end{subfigure}%
	\begin{subfigure}{0.5\textwidth}
		\centering
		\input{CovRadius_Power.tex}
		\caption{}
		\label{CovRadius}
	\end{subfigure}
	\caption{(a) For a target zone indicated by its radius, the range of sensing powers are influenced significantly by the altitude of flying SDrs. (b) For an assumed minimum ADr's transmit power an SDr can fly at an optimum altitude to extend its monitoring region compared to ground surveillance.}
	\label{}
 	\vspace{-0.4cm}
\end{figure*}
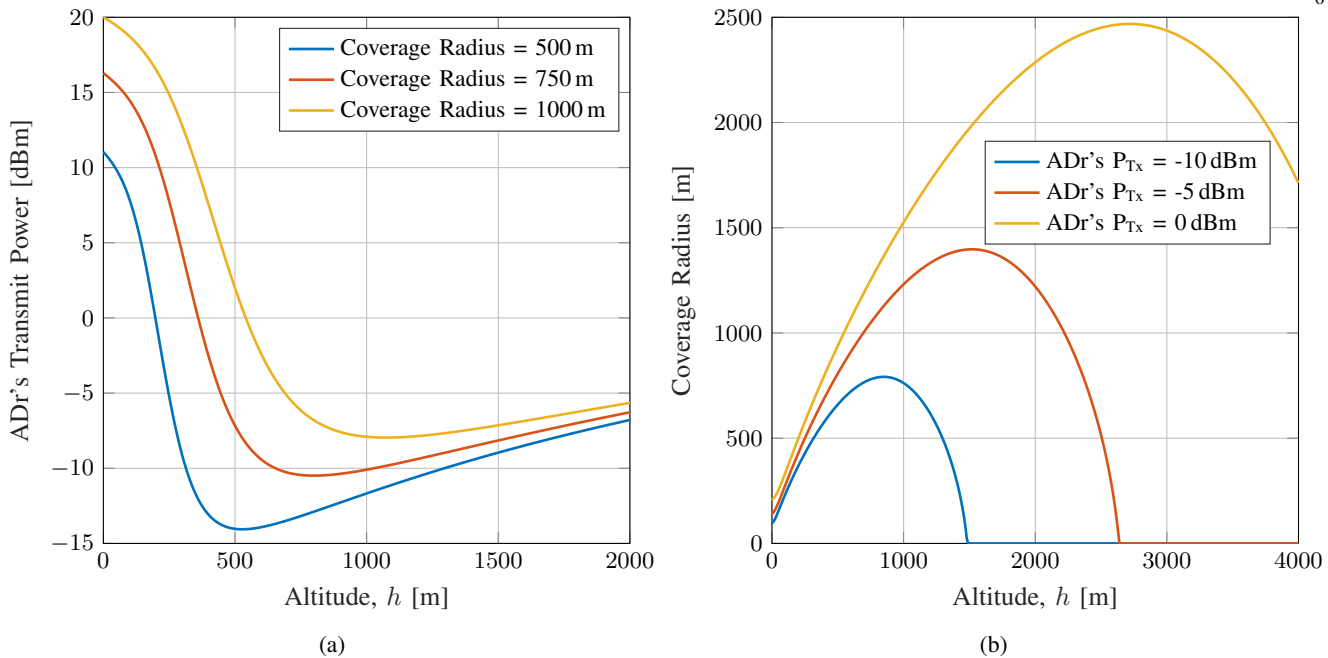

\subsection{A Target Control Zone} \label{optimization-power}

Considering a target control zone for an SDr flying at altitude $h$, the fact that the ADr's transmit power is unknown means that several flight levels have to be defined based on the range of possible transmit powers rather than a single optimal altitude. In any case, in order to detect the drone, it should be within the coverage of the SDr. Figure \ref{PowGain} illustrates a lower bound of the range of ADr's transmit power that is detectable by an SDr at each altitude. This lower bound is obtained by comparing the received SNR and a certain threshold. For instance, if the SDr flies at 200\,m above a target region of radius 500\,m, only an ADr transmitting higher than 0\,dBm can be sensed within the whole target zone. This figure also proposes an optimum altitude at which the lowest possible transmit power is detectable. As a matter of fact, if the SDr goes higher the LoS probability increases resulting in better channel quality as explained in the previous section, however the link length also increases deteriorating the channel quality due to path loss. These opposite effects are balanced at the optimum altitude shown in the figure. From this figure it can be seen that, as the target region becomes larger the minimum required transmit power of ADrs and the optimum altitude of SDrs increases. Please note that if the ADr flies higher over the ground, then the link to the SDr will become LoS and hence the channel quality will increase resulting in a lower detectable power and a higher classification accuracy of the technology used in the ADr.

\subsection{Coverage Extension} \label{optimization-coverage}

In this subsection we present an efficient deployment of an SDr to maximize the covered region assuming a minimum ADr P$_{\text{Tx}}$. Therefore, the result can be used to find the optimal number of surveillance drones in order to cover a larger target region. Figure \ref{CovRadius} illustrates the impact of altitude for different minimum transmit powers. The figure shows that as the drone goes higher the coverage increases such that at an optimum altitude the coverage is maximized. For instance, an SDr can fly at an altitude of 900\,m to maximize the region of control assuming a minimum ADr's transmit power of -10\,dBm. More technical discussion for the optimal deployment of the drones and the trade-offs can be found in \cite{azari2016joint,azari2017ultra,mozaffari2016efficient}.

\subsection{Localizing Amateur Drones} \label{optimization-localization}

Once the ADr has been detected and its RSS has been measured, we can proceed to estimate its location. In our simulation we assume 3 SDrs positioned as vertices of an equilateral triangle with sides of length $l$ and equal adjustable altitude $h$. Each SDr has its own coverage radius and is able to detect any ADr separately. The intersection of the three coverage areas produces the \textit{no-fly zone} associated with the three SDrs combined.

\begin{figure*}[t]
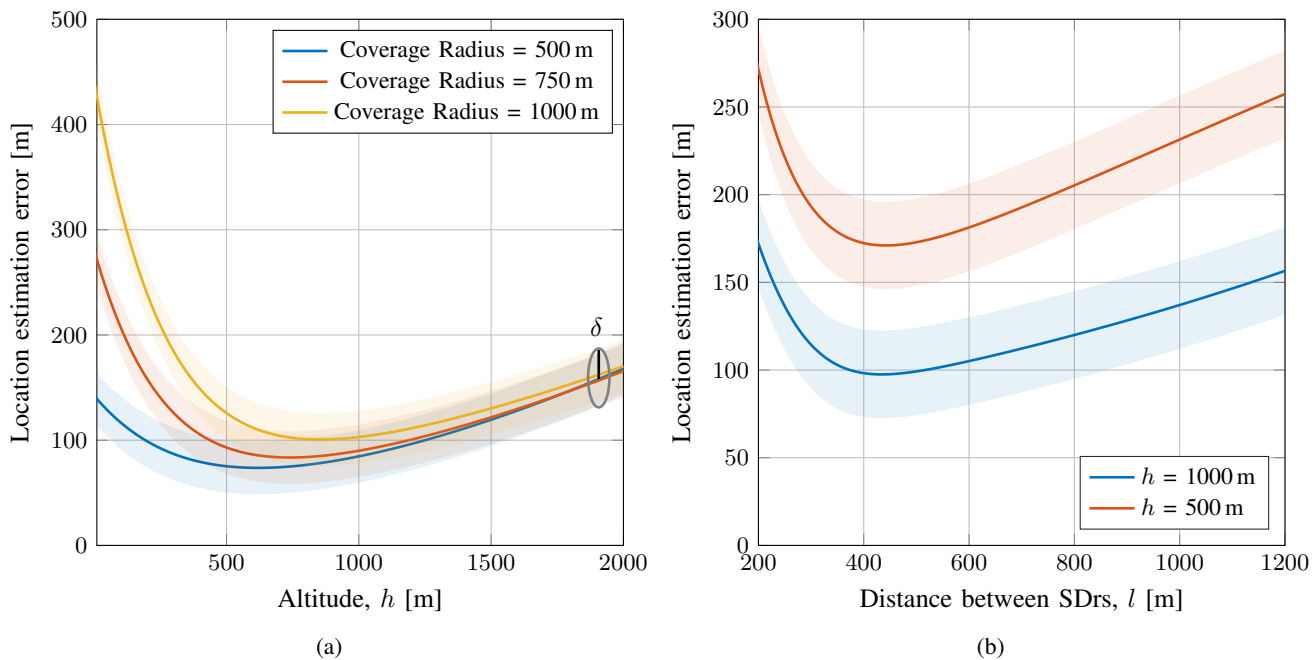

\centering
	\begin{subfigure}{.5\textwidth}
		\centering
		\input{estD.tex}
		\caption{}
		\label{errh}
	\end{subfigure}%
	\begin{subfigure}{.5\textwidth}
		\centering
		\input{estDL.tex}
		\caption{}
		\label{errl}
	\end{subfigure}
	\caption{Localization of ADrs is affected by the altitude $h$ and the distance $l$ between SDrs: (a) Localization error as a function of the SDrs altitude with the uncertainty constant $\delta$. (b) Localization error as function of distance $l$ between SDrs.}
	\label{LocError}
 	\vspace{-0.4cm}
\end{figure*}

Unlike the G2G RSS-based localization scenarios which suffer from shadowing as the main source of error, introducing the altitude in G2A and A2A scenarios as a third dimension promises to overcome the shadowing due to the high $\plos$ and hence minimize the localization error. In this subsection, we present the localization of ADrs by using SDrs as shown in Figure~\ref{loca}. As illustrated in the figure, we target localizing any ADr that would fly in the defined \textit{no-fly zone} by means of the RSS-based distance estimation. In order to define the position of the ADr, distances to three different SDrs need to be estimated using RSS. However, as the transmit power from the ADr is unknown, one can define a constant $\text{P}^{(min)}_{\text{Tx}} \leq C \leq \text{P}^{(max)}_{\text{Tx}}$ that represents the possible transmit power. Accordingly, the estimated distance between the ADr and the SDr is equal to $\hat{d} + \delta$ where $\delta$ is a constant that represents the uncertainty due to the unknown transmit power. Subsequently, after estimating the distance between the ADr and three SDrs, we will end up with a sphere of radius $\delta$ in which the ADr is located. A representation of $\delta$ is shown in Figure~\ref{errh} which can cause under- or over-estimation of the distance.

As shown in previous sections, the altitude of the drones has a significant influence on the model representing the received power and hence, the accuracy of the estimated location of the ADr. In fact, for any SDr, the variations of both the the path loss exponent and shadowing standard deviation with the altitude are modeled based on statistical representations given in equations (\ref{plexpo}) and (\ref{shdow}), respectively. The shadowing effect at low values of $h$ will be relatively high causing large localization errors. As $h$ increases, the shadowing effect will decrease, concurrently, the resolution\footnote{The resolution is the ability to distinguish two different distances from two different RSS measurements} will also decrease. In the case of low resolution, any small variation will bring a large localization error. Therefore, based on the behavior of the path loss model, the existence of an optimal altitude is investigated as shown in Figure~\ref{LocError}. Our results show that for a targeted \textit{no-fly zone}, an optimum altitude $h$ minimizing the localization error is present. This optimal altitude is shown in Figure~\ref{errh}. As one can see, an optimal altitude for minimizing the estimated location error exists at $h$~=~800\,m for a coverage radius of 1000\,m. Moreover, it can be seen that when the SDrs are at the low altitude of $< 800$\,m, the RSS is exposed to high shadowing resulting a relatively high estimation error assuming the same coverage. On the other hand, high altitude means low resolution in the RSS-distance curve where any low shadowing causes a relativity high estimation error. It is worth noting that, the coverage radius here can be connected to the P$_{\text{Tx}}$, since the lower the P$_{\text{Tx}}$ of ADr, the smaller the coverage radius of the SDr at which ADrs are detectable.

In addition to the altitude dependence, the localization accuracy is affected by the distance $l$ between the SDrs. Figure \ref{errl} illustrates the location estimation error against the distances $l$ under an assumption of the equidistant SDrs positioning. As shown in the figure, when $l$ is relatively small, the intersection zone between the estimated distances of the three SDrs will be large leading to a relatively high localization error. Moreover, increasing $l$ improves the localization accuracy where an accuracy of 100\,m is achieved at $l$~=~300\,m for the same $h$ of 1000\,m. Eventually, as $l$ keeps increasing, the distance will become too large to give an acceptable resolution for distance estimate, making a low localization error impossible.

\section{Research Directions and Future Work} \label{future}

Gradually, drones are gaining a lot of momentum as an ingrained part of future wireless technologies making surveillance of amateur drones very important. In this section we address some open problems as future work towards auxiliary reliable surveillance of ADrs.
\subsection{Experimental channel models validation}
Although, some interesting works have been carried out to model the G2A channel characteristic using measurements or simulations, they only considered either non-urban environments or relatively high altitudes. In fact, a generic channel model that reflects characteristics of both A2A and G2A channels and dependency on the elevation angle including the shadowing effect is required. To this end, a measurement campaign in order to validate the different proposed simulation-based models and define a generic one is still one of the future plans.
\subsection{Tracking of ADrs}
In this work we consider localizing the ADrs at a given time, however, as the ADrs are mobile in nature, localization must be considered over time in order to track the ADrs' movement. To this end, the processing time between detecting the ADr, defined a valid RSS measurement and estimating its position need to be minimized. In fact, this time has to be lower than the time needed for the ADr to move a certain distance that defined the required localization accuracy. 

\subsection{RF fingerprinting for ADr identification}

Even though we have explained various state-of-the-art techniques for ADr detection and technology classification, identifying a drone with passive RF monitoring is quite challenging. For detecting ADrs, temporal and spatial wireless transmission statistics should be derived from the received detected signal which should be further associated with a particular drone. Detailed studies should be done to enable and improve RF fingerprinting for drone surveillance. We believe drone identification with low false identification rates can be achieved by combining RF localization and fingerprinting.

\subsection{Mobility aided surveillance}
In order to decrease the number of surveillance drones in a given region, a mobile drone can be deployed. A mobile drone will not only increase the number of accessible ground station but it can also localize other drones by collecting measurements at different locations. However, the cost to be payed is the delay although the speed and the trajectory of movement can be optimized for the minimum delay.

\subsection{Network lifetime}
In order to increase network lifetime we need to optimize the power consumption. Since mechanical power is the main source of the energy cost, power consumption can be improved by optimizing the flying trajectory. Considering the trajectory of movement, the duration of communication, the payload weight and the battery size, the lifetime of a drone is limited to less than one hour. Therefore, an alternative solutions such as solar cells (e.g., Facebook Aquila Drone) for providing the required mechanical energy is of high importance. Nevertheless, adding solar cells also increases the rate of energy consumption as more weight has to be carried on the drone.

\section{Conclusion}
In this article we have considered a network of SDrs aiming to sense the presence of ADrs and localize them over a \textit{no-fly zone} by means of the passive RF sensing. An overview of the state-of-the-art RF passive sensing and detection showed that an SNR above 0\,dB is required for accurate detection of ADrs. However, using aerial based sensors improves the received SNR due to better channels between ADrs flying at low altitudes and SDrs. Therefore, the characteristics of the channel between SDrs and ADrs have been thoroughly studied considering the worst case scenario at which the ADrs are 2\,m above the ground. Our results show that a tenfold increase of the coverage radius and a 25\,dB reduction of the minimum detectable power can be achieved by flying the SDrs at the optimal altitude. Furthermore, it has been shown that 4 times better localization accuracy is gained by careful optimizing the altitude of the SDrs for a given \textit{no-fly zone}. We expect that academic and industrial research and development activities can use the proposed framework to address the drone surveillance challenges introduced in the paper.

\bibliographystyle{IEEEtran}
\bibliography{bibliography}

\begin{IEEEbiography}
[{\includegraphics[width=1in,height=1.25in]{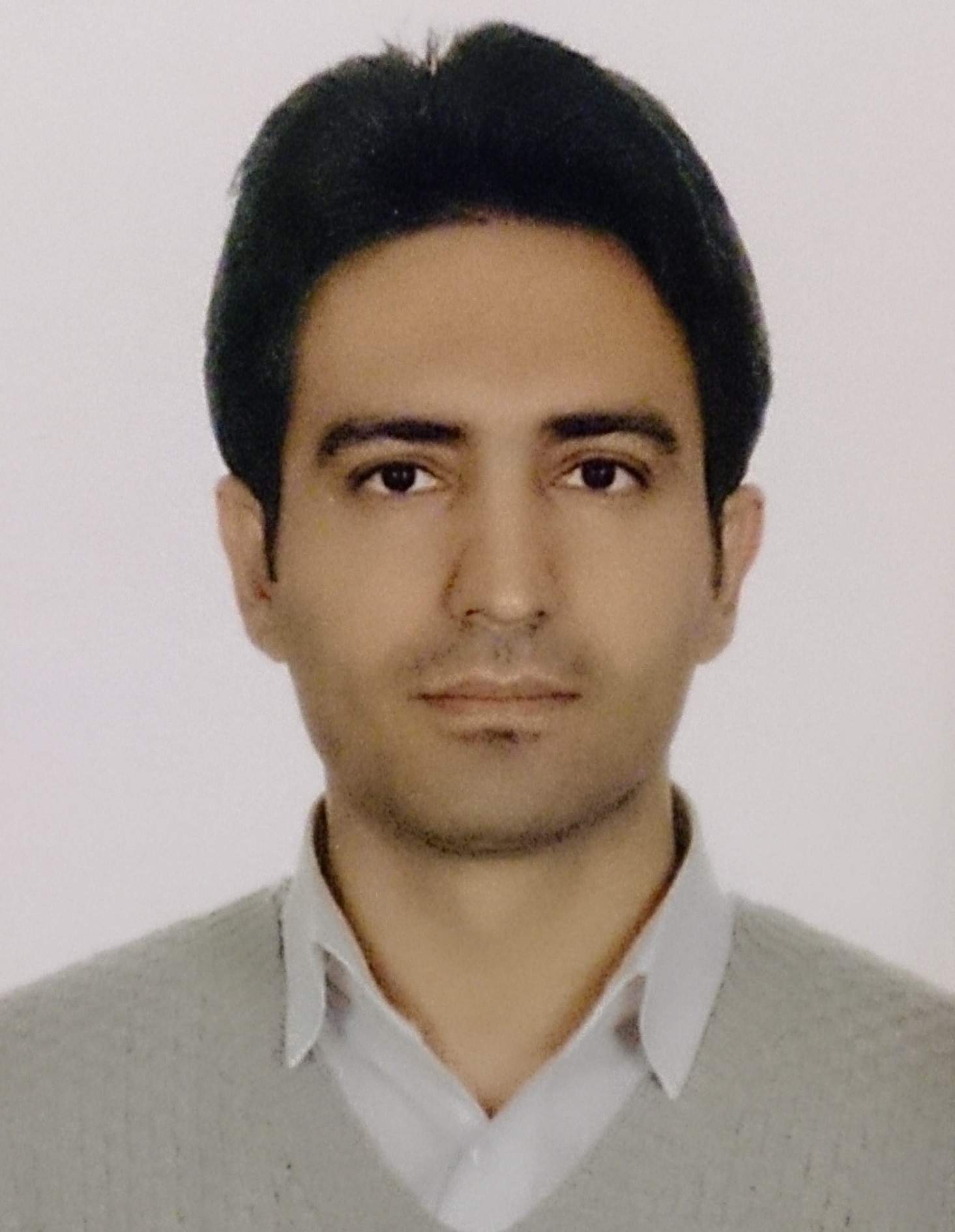}}]
{Mohammad Mahdi Azari\,\,\,\,\,\,\,\,\,\,\,\,\,\,\,\,\,\,\,\,\,\,\,\,\,\,\,\,\,\,\,\,\,\,\,\,\,\,\,\,\,\,\,\,\,\,\,\,\,\,}
(mmahdi.azari@gmail.com) has received the B.Sc. and  M.Sc. degrees in electrical and communication engineering from University of Tehran, Tehran, Iran. Currently he is a Ph.D. candidate at the Department of Electrical Engineering, KU Leuven, Belgium. His main research interests include unmanned aerial vehicle (UAV) communication and networking, modeling and analysis of cellular networks, and mmWave communication. He has also received Iran's National Elites Foundation (INEF) Award.\end{IEEEbiography}

\begin{IEEEbiography}
[{\includegraphics[width=1in,height=1.25in]{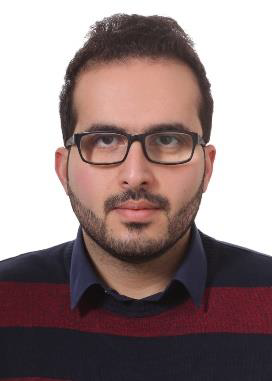}}]
{Hazem Sallouha}(hazem.sallouha@kuleuven.be) received the B.Sc. degree in electrical engineering from Islamic University of Gaza, Palestine, in 2011, the M.Sc. degree in electrical engineering majoring in wireless communications from Jordan University of Science and Technology, Jordan in 2013. Currently he is a Ph.D. candidate at the Department of Electrical Engineering, KU Leuven, Belgium. His main research interests include localization techniques, communications with unmanned aerial vehicles, Internet of things networks and machine learning algorithms for localization.\end{IEEEbiography}

\begin{IEEEbiography}
[{\includegraphics[width=1in,height=1.25in]{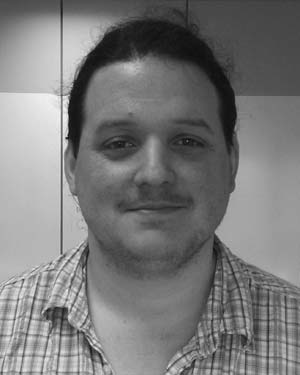}}]
{Alessandro Chiumento \,\,\,\,\,\,\,\,\,\,\,\,\,\,\,\,\,\,\,\,\,\,\,\,\,\,\,\,\,\,\,\,\,\,\,\,\,\,\,\,\,\,\,\,\,\,\,\,\,\,\,\,\,\,}(alessandro.chiumento@esat.kuleuven.be) received his Ph.D. degree in cellular network management from Imec, Leuven, Belgium, in 2015. He is currently with the Department of Electrical Engineering, Katholieke Universiteit Leuven. His research interests include massive machine-to-machine communication, channel prediction, very dense networks, and the application of machine learning to theoretical problems in telecommunication and information management.\end{IEEEbiography}

\begin{IEEEbiography}
[{\includegraphics[width=1in,height=1.25in]{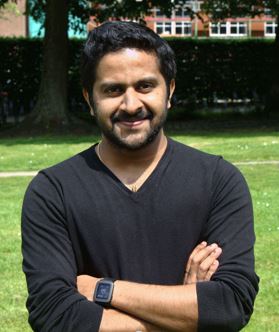}}]{Sreeraj Rajendran} received his Masters degree in communication and signal processing from the Indian Institute of Technology, Bombay, in 2013. He is currently pursuing the Ph.D. degree in the Department of Electrical Engineering, KU Leuven. Before joining KU Leuven, he worked as a senior design engineer in the baseband team of Cadence and ASIC verifidetecation engineer in Wipro Technologies. His main research interests include machine learning algorithms for wireless and low power wireless sensor networks.\end{IEEEbiography}

\balance

\begin{IEEEbiography}[{\includegraphics[width=1in,height=1.25in]{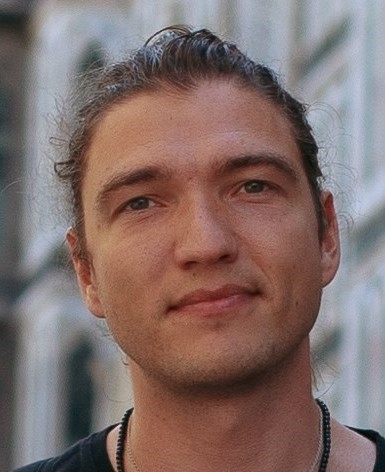}}]{Evgenii Vinogradov} received the Dipl. Engineer degree in Radio Engineering and Telecommunications from Saint-Petersburg Electrotechnical University (Russia), in 2009. After several years of working in the field of mobile communications, he joined UCL (Belgium) in 2013, where he obtained his Ph.D. degree in 2017. His doctoral research interests focused on radio propagation channel modeling. In 2017, Evgenii joined the electrical engineering department at KU Leuven (Belgium) where he is working on wireless communications with UAVs.\end{IEEEbiography}

\begin{IEEEbiography}[{\includegraphics[width=1in,height=1.25in]{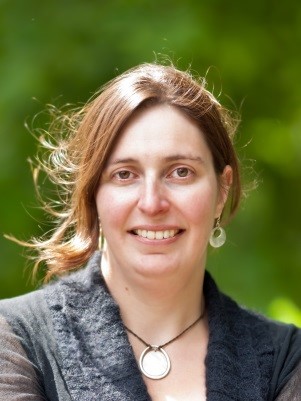}}]{Sofie Pollin} Sofie Pollin obtained her PhD at KU Leuven in 2006. She continued her research on wireless communication at UC Berkeley.   In November 2008 she returned to imec to become a principal scientist in the green radio team. Since 2012, she is tenure track assistant professor at the electrical engineering department at KU Leuven. Her research centers around Networked Systems that require networks that are ever more dense, heterogeneous, battery powered and spectrum constrained. \end{IEEEbiography}

\end{document}

%% file: PL_tet_alpha.tex
%
%
\pgfplotsset{every tick label/.append style={font=\small}}

\definecolor{mycolor1}{rgb}{0.00000,0.44700,0.74100}%
\definecolor{mycolor2}{rgb}{0.85000,0.32500,0.09800}%
\begin{tikzpicture}

\begin{axis}[%
legend style={font=\fontsize{9}{5}\selectfont}, 
width=6.2cm,
height=7cm,
at={(0.758in,0.481in)},
scale only axis,
xmin=0,
xmax=60,
xlabel style={font=\color{white!15!black}},
xlabel={Elevation Angle, $\theta~[^o]$},
xtick={0,10,20,30,40,50,60},
separate axis lines,
every outer y axis line/.append style={mycolor1},
every y tick label/.append style={font=\small\color{mycolor1}},
every y tick/.append style={mycolor1},
ymin=2,
ymax=3,
ytick={  2, 2.2, 2.4, 2.6, 2.8,   3},
ylabel style={font=\color{mycolor1}},
ylabel={Path Loss Exponent, $\alpha(\theta)$},
axis background/.style={fill=white},
xmajorgrids,
ymajorgrids
]
\addplot [color=mycolor1,line width=1.0pt]
  table[row sep=crcr]{%
0	2.97774301279498\\
0.666666666666667	2.97533007053999\\
1.33333333333333	2.97266284848196\\
2	2.96971621001709\\
2.66666666666667	2.96646290820522\\
3.33333333333333	2.96287349406262\\
4	2.95891624220256\\
4.66666666666667	2.9545571002571\\
5.33333333333333	2.94975966968354\\
6	2.94448522679942\\
6.66666666666667	2.93869279415611\\
7.33333333333333	2.93233927358848\\
8	2.92537965337485\\
8.66666666666667	2.91776730278695\\
9.33333333333333	2.90945436774863\\
10	2.90039228116584\\
10.6666666666667	2.89053240052062\\
11.3333333333333	2.87982678329363\\
12	2.86822910744689\\
12.6666666666667	2.85569573932618\\
13.3333333333333	2.84218694475505\\
14	2.82766823070617\\
14.6666666666667	2.81211179482335\\
15.3333333333333	2.79549804850403\\
16	2.77781716677382\\
16.6666666666667	2.75907060561792\\
17.3333333333333	2.73927251589923\\
18	2.71845097385919\\
18.6666666666667	2.69664894298093\\
19.3333333333333	2.67392488219217\\
20	2.65035292225525\\
20.6666666666667	2.62602254650332\\
21.3333333333333	2.60103773387243\\
22	2.57551555057919\\
22.6666666666667	2.54958420995751\\
23.3333333333333	2.52338065517002\\
24	2.49704775340904\\
24.6666666666667	2.47073121925433\\
25.3333333333333	2.44457640581981\\
26	2.41872511275626\\
26.6666666666667	2.39331255883985\\
27.3333333333333	2.36846465392588\\
28	2.34429568202256\\
28.6666666666667	2.3209064767999\\
29.3333333333333	2.29838313634977\\
30	2.27679628899291\\
30.6666666666667	2.25620088962417\\
31.3333333333333	2.236636499042\\
32	2.21812797855194\\
32.6666666666667	2.20068651952253\\
33.3333333333333	2.18431092228544\\
34	2.16898903990862\\
34.6666666666667	2.15469930859951\\
35.3333333333333	2.14141229629598\\
36	2.12909221288021\\
36.6666666666667	2.11769833808509\\
37.3333333333333	2.10718633549704\\
38	2.09750943233417\\
38.6666666666667	2.08861945442036\\
39.3333333333333	2.08046771375863\\
40	2.07300575229916\\
40.6666666666667	2.06618595000647\\
41.3333333333333	2.05996200834965\\
42	2.05428932210854\\
42.6666666666667	2.04912525315621\\
43.3333333333333	2.04442931988246\\
44	2.04016331537785\\
44.6666666666667	2.03629136658399\\
45.3333333333333	2.0327799454799\\
46	2.02959784213174\\
46.6666666666667	2.0267161081674\\
47.3333333333333	2.02410797801027\\
48	2.02174877405572\\
48.6666666666667	2.01961580092522\\
49.3333333333333	2.01768823299649\\
50	2.01594699858806\\
50.6666666666667	2.014374663469\\
51.3333333333333	2.01295531576355\\
52	2.0116744538155\\
52.6666666666667	2.01051887815935\\
53.3333333333333	2.00947658840304\\
54	2.00853668555049\\
54.6666666666667	2.00768928007173\\
55.3333333333333	2.00692540585523\\
56	2.00623694004297\\
56.6666666666667	2.00561652864776\\
57.3333333333333	2.00505751777723\\
58	2.00455389023608\\
58.6666666666667	2.00410020724192\\
59.3333333333333	2.00369155496793\\
60	2.00332349561362\\
};
\label{plotyyref:leg1}

\end{axis}

\begin{axis}[%
legend style={font=\fontsize{9}{5}\selectfont}, 
width=6.2cm,
height=7cm,
at={(0.758in,0.481in)},
scale only axis,
xmin=0,
xmax=60,
xtick={0,10,20,30,40,50,60},
every outer y axis line/.append style={mycolor2},
every y tick label/.append style={font=\small\color{mycolor2}},
every y tick/.append style={mycolor2},
ymin=0,
ymax=1,
ytick={  0, 0.2, 0.4, 0.6, 0.8,   1},
ylabel style={font=\color{mycolor2}},
ylabel={LoS Probability, $\mathcal{P}_\mathrm{LoS}(\theta)$},
axis y line*=right,
legend style={at={(0.634,0.10)}, anchor=south west, legend cell align=left, align=left, draw=white!15!black}
]
\addlegendimage{/pgfplots/refstyle=plotyyref:leg1}
\addlegendentry{$\alpha(\theta)$}
\addplot [color=mycolor2,line width=1.0pt]
  table[row sep=crcr]{%
0	0.0222569872050231\\
0.666666666666667	0.0246699294600152\\
1.33333333333333	0.0273371515180401\\
2	0.0302837899829062\\
2.66666666666667	0.0335370917947778\\
3.33333333333333	0.037126505937384\\
4	0.0410837577974401\\
4.66666666666667	0.0454428997428973\\
5.33333333333333	0.0502403303164567\\
6	0.0555147732005826\\
6.66666666666667	0.0613072058438888\\
7.33333333333333	0.0676607264115163\\
8	0.0746203466251515\\
8.66666666666667	0.0822326972130454\\
9.33333333333333	0.0905456322513746\\
10	0.0996077188341577\\
10.6666666666667	0.10946759947938\\
11.3333333333333	0.120173216706368\\
12	0.131770892553115\\
12.6666666666667	0.144304260673818\\
13.3333333333333	0.157813055244949\\
14	0.172331769293834\\
14.6666666666667	0.187888205176654\\
15.3333333333333	0.204501951495971\\
16	0.22218283322618\\
16.6666666666667	0.240929394382078\\
17.3333333333333	0.260727484100768\\
18	0.281549026140811\\
18.6666666666667	0.303351057019066\\
19.3333333333333	0.326075117807829\\
20	0.34964707774475\\
20.6666666666667	0.373977453496679\\
21.3333333333333	0.398962266127572\\
22	0.424484449420811\\
22.6666666666667	0.450415790042493\\
23.3333333333333	0.476619344829977\\
24	0.502952246590965\\
24.6666666666667	0.529268780745672\\
25.3333333333333	0.55542359418019\\
26	0.581274887243739\\
26.6666666666667	0.606687441160154\\
27.3333333333333	0.631535346074118\\
28	0.655704317977443\\
28.6666666666667	0.679093523200099\\
29.3333333333333	0.701616863650231\\
30	0.723203711007093\\
30.6666666666667	0.743799110375833\\
31.3333333333333	0.763363500957997\\
32	0.781872021448058\\
32.6666666666667	0.799313480477472\\
33.3333333333333	0.815689077714564\\
34	0.831010960091382\\
34.6666666666667	0.845300691400488\\
35.3333333333333	0.858587703704018\\
36	0.870907787119789\\
36.6666666666667	0.882301661914913\\
37.3333333333333	0.892813664502963\\
38	0.902490567665832\\
38.6666666666667	0.911380545579637\\
39.3333333333333	0.919532286241373\\
40	0.926994247700843\\
40.6666666666667	0.93381404999353\\
41.3333333333333	0.94003799165035\\
42	0.945710677891458\\
42.6666666666667	0.950874746843786\\
43.3333333333333	0.955570680117542\\
44	0.959836684622149\\
44.6666666666667	0.963708633416012\\
45.3333333333333	0.967220054520099\\
46	0.970402157868265\\
46.6666666666667	0.973283891832597\\
47.3333333333333	0.975892021989725\\
48	0.978251225944276\\
48.6666666666667	0.980384199074783\\
49.3333333333333	0.982311767003512\\
50	0.984053001411943\\
50.6666666666667	0.985625336530998\\
51.3333333333333	0.987044684236447\\
52	0.988325546184501\\
52.6666666666667	0.989481121840651\\
53.3333333333333	0.990523411596956\\
54	0.991463314449506\\
54.6666666666667	0.992310719928265\\
55.3333333333333	0.993074594144771\\
56	0.993763059957027\\
56.6666666666667	0.994383471352237\\
57.3333333333333	0.994942482222768\\
58	0.995446109763922\\
58.6666666666667	0.995899792758077\\
59.3333333333333	0.996308445032073\\
60	0.996676504386382\\
};
\addlegendentry{$\mathcal{P}_\mathrm{LoS}(\theta)$}

\end{axis}
\end{tikzpicture}%

%% file: Power_CovRadius.tex
%
%
\pgfplotsset{every tick label/.append style={font=\small}}
\definecolor{mycolor1}{rgb}{0.00000,0.44700,0.74100}%
\definecolor{mycolor2}{rgb}{0.85000,0.32500,0.09800}%
\definecolor{mycolor3}{rgb}{0.92900,0.69400,0.12500}%
\begin{tikzpicture}

\begin{axis}[%
legend style={font=\fontsize{9}{5}\selectfont},
width=7cm,
height=7cm,
at={(0.974in,0.632in)},
scale only axis,
xmin=0,
xmax=2000,
xlabel style={font=\color{white!15!black}},
xlabel={Altitude, $h$ [m]},
ymin=-15,
ymax=20,
ylabel style={font=\color{white!15!black}},
ylabel={ADr's Transmit Power [dBm]},
axis background/.style={fill=white},
xmajorgrids,
ymajorgrids,
legend style={legend cell align=left, align=left, draw=white!15!black}
]
\addplot [color=mycolor1,line width=1.0pt]
  table[row sep=crcr]{%
0	11.038021275984\\
2.33490560374371	10.992885195032\\
4.66991304491963	10.9470192284751\\
7.0051241787274	10.9003830944293\\
9.34064089590808	10.8529351518796\\
11.6765651405314	10.8046322943479\\
14.0129989278026	10.7554298210651\\
16.3500443618961	10.7052817786423\\
18.6878036538213	10.6541404764961\\
21.0263791393294	10.6019567079575\\
23.3658732968651	10.5486796750674\\
25.7063887655726	10.4942570013361\\
28.0480283633606	10.4386345256147\\
30.390895105034	10.3817566582516\\
32.7350922204994	10.3235659775454\\
35.0807231730504	10.2640033918564\\
37.4278916777403	10.2030082128356\\
39.776701719849	10.1405178798649\\
42.1272575734507	10.0764683512489\\
44.4796638200906	10.0107936758571\\
46.8340253675761	9.94342638704028\\
49.1904474688918	9.87429725601893\\
51.5490357412437	9.80333547731632\\
53.9098961852409	9.73046843508374\\
56.273135204223	9.65562222142191\\
58.6388596237387	9.57872118386686\\
61.0071767111861	9.49968826473443\\
63.3781941956199	9.41844492876515\\
65.7520202877351	9.33491134775667\\
68.1287637000344	9.2490063937165\\
70.5085336671876	9.16064776710341\\
72.891439966592	9.06975211052016\\
75.2775929391407	8.97623509084866\\
77.6671035102087	8.88001154471333\\
80.0600832108653	8.7809957279633\\
82.4566441993205	8.67910119953524\\
84.8568992826158	8.57424139974235\\
87.2609619385682	8.46632937445195\\
89.6689463379766	8.35527847989306\\
92.0809673670986	8.24100203509769\\
94.4971406504114	8.12341416460656\\
96.917582573662	8.00242959716897\\
99.3424103072197	7.87796412455654\\
101.771741829741	7.74993486646824\\
104.205695952155	7.6182605734477\\
106.644392341987	7.48286194718784\\
109.087951548016	7.34366198363959\\
111.536495025301	7.20058632778068\\
113.990145160561	7.05356362920556\\
116.449025297944	6.90252596131613\\
118.91325976518	6.7474092042103\\
121.382973900139	6.58815340083818\\
123.858294077803	6.4247032540749\\
126.339347737671	6.25700862166055\\
128.826263411595	6.08502470272848\\
131.319170752086	5.90871274206138\\
133.818200561075	5.72804031241363\\
136.323484819161	5.54298197113042\\
138.835156715364	5.3535193726749\\
141.353350677375	5.15964195633948\\
143.878202402351	4.96134737618682\\
146.409848888241	4.75864162212875\\
148.948428465674	4.5515398968067\\
151.494080830428	4.34006660030531\\
154.046947076485	4.124255730155\\
156.6071697297	3.90415144576286\\
159.1748927821	3.67980813635066\\
161.750261726824	3.45129063991088\\
164.333423593735	3.21867460862427\\
166.924526985711	2.98204675630197\\
169.523722115645	2.7415047517444\\
172.131160844171	2.4971571944165\\
174.74699671813	2.24912421880484\\
177.371385009814	1.99753673491884\\
180.004482756989	1.74253675482467\\
182.646448803746	1.48427717578937\\
185.297443842175	1.22292151718005\\
187.957630454908	0.958643616787938\\
190.627173158549	0.691627208403759\\
193.306238448015	0.422065662766542\\
195.994994841812	0.150161544905458\\
198.693612928286	-0.123874072653976\\
201.402265412862	-0.39982234301398\\
204.121127166308	-0.677456754784757\\
206.850375274055	-0.956544392660959\\
209.590189086597	-1.23684641154269\\
212.340750271014	-1.51811886098683\\
215.10224286364	-1.80011353672725\\
217.874853323913	-2.08257877538249\\
220.658770589445	-2.36526052315915\\
223.45418613235	-2.64790292754057\\
226.261294016855	-2.93024941734257\\
229.080290958246	-3.21204383702462\\
231.911376383187	-3.49303084736411\\
234.754752491439	-3.77295733855306\\
237.61062431904	-4.05157291169202\\
240.47919980298	-4.32863109686352\\
243.360689847414	-4.60388997039516\\
246.255308391463	-4.87711294582769\\
249.163272478657	-5.14806965976673\\
252.084802328052	-5.41653661539711\\
255.020121407097	-5.68229782420671\\
257.969456506281	-5.94514558594452\\
260.933037815628	-6.20488071559339\\
263.911099003096	-6.46131334772788\\
266.903877294932	-6.71426332257926\\
269.911613558052	-6.96356036465782\\
272.934552384499	-7.20904460938064\\
275.972942178059	-7.45056679032848\\
279.027035243081	-7.68798832408584\\
282.097087875596	-7.92118165410295\\
285.18336045678	-8.15003002029771\\
288.28611754886	-8.37442774760123\\
291.405627993525	-8.59427998380174\\
294.542165012924	-8.80950283455134\\
297.696006313343	-9.0200230112977\\
300.867434191633	-9.22577758161071\\
304.056735644493	-9.42671409884044\\
307.264202480684	-9.62278997405059\\
310.490131436285	-9.81397230289788\\
313.734824293081	-10.000237525039\\
316.998588000184	-10.1815710850878\\
320.281734799009	-10.3579671148879\\
323.58458235169	-10.5294278663561\\
326.907453873073	-10.6959634410888\\
330.250678266399	-10.8575912806745\\
333.614590262795	-11.0143358080896\\
336.999530564702	-11.1662280329125\\
340.405845993385	-11.3133049195285\\
343.833889640647	-11.4556091668292\\
347.284021024902	-11.593188663807\\
350.756606251748	-11.726096121265\\
354.25201817921	-11.854388576124\\
357.770636587789	-11.9781270996156\\
361.312848355512	-12.0973762749609\\
364.879047638139	-12.2122038993987\\
368.469636054712	-12.3226806212655\\
372.085022878641	-12.4288795081872\\
375.725625234516	-12.5308757761904\\
379.391868300844	-12.6287464322711\\
383.084185518946	-12.7225699853495\\
386.8030188082	-12.8124261422198\\
390.548818787894	-12.898395518641\\
394.322045005902	-12.9805594658882\\
398.123166174457	-13.0589996372852\\
401.952660413249	-13.1337980439219\\
405.811015500156	-13.2050365347802\\
409.69872912986	-13.2727968436528\\
413.616309180655	-13.3371602965872\\
417.564273989753	-13.3982076779029\\
421.543152637409	-13.456019067607\\
425.553485240196	-13.5106737057558\\
429.595823253771	-13.5622498785047\\
433.670729785512	-13.6108247587657\\
437.778779917396	-13.6564745020948\\
441.920561039502	-13.6992737598975\\
446.096673194578	-13.7392960187447\\
450.307729434081	-13.7766133460608\\
454.554356186155	-13.8112962459752\\
458.83719363601	-13.8434137768152\\
463.156896119199	-13.8730334490735\\
467.514132528311	-13.9002211243929\\
471.909586733611	-13.9250410748854\\
476.343958018203	-13.9475558815731\\
480.817961528287	-13.9678266066563\\
485.332328739153	-13.9859124404708\\
489.887807937534	-14.0018710122704\\
494.485164721012	-14.015758259019\\
499.12518251518	-14.0276284024285\\
503.808663109297	-14.0375341156884\\
508.536427211231	-14.0455262795032\\
513.309315022488	-14.0516541578355\\
518.128186834199	-14.0559654383748\\
522.99392364495	-14.0585062356905\\
527.907427801406	-14.059320892399\\
532.869623662721	-14.0584524370521\\
537.881458289757	-14.0559421244983\\
542.94390216022	-14.0518298836796\\
548.057949910849	-14.0461540122929\\
553.224621107867	-14.0389514230224\\
558.444961046954	-14.0302575421594\\
563.720041584073	-14.0201064834954\\
569.05096199856	-14.0085309439173\\
574.43884988993	-13.9955621488575\\
579.884862109972	-13.9812302154295\\
585.390185731744	-13.9655638074056\\
590.956039057199	-13.948590407532\\
596.583672665254	-13.9303362517776\\
602.2743705022	-13.9108263968173\\
608.029451016469	-13.8900847706349\\
613.850268339891	-13.8681339824814\\
619.738213517659	-13.8449957233856\\
625.694715789381	-13.820690480045\\
631.721243923708	-13.7952377734071\\
637.81930760917	-13.7686560208541\\
643.990458904006	-13.7409626286605\\
650.236293747934	-13.7121740362387\\
656.558453538961	-13.6823057413678\\
662.958626778539	-13.6513723170402\\
669.438550788537	-13.6193873946289\\
676.000013503719	-13.5863637053656\\
682.644855343644	-13.5523131134633\\
689.3749711681	-13.5172466233569\\
696.19231232049	-13.4811745556864\\
703.098888763789	-13.4441061793149\\
710.096771314029	-13.4060501828402\\
717.188093976537	-13.3670143164738\\
724.375056390486	-13.3270056761814\\
731.659926387671	-13.2860305823928\\
739.045042671781	-13.2440946273078\\
746.532817624851	-13.2012026895035\\
754.125740247982	-13.1573589455529\\
761.826379243898	-13.1125668819277\\
769.637386249385	-13.0668292691247\\
777.561499226178	-13.0201483205893\\
785.60154601945	-12.9725255201523\\
793.760448093624	-12.9239616275295\\
802.041224455948	-12.8744568375459\\
810.446995778881	-12.8240107890827\\
818.980988733205	-12.7726223286174\\
827.64654054449	-12.7202897808741\\
836.447103786496	-12.6670108339292\\
845.386251425999	-12.6127825565909\\
854.467682134564	-12.5576014021505\\
863.695225883903	-12.5014631972814\\
873.072849842628	-12.4443632332224\\
882.604664593537	-12.3862960777711\\
892.294930691908	-12.3272557328571\\
902.148065586882	-12.2672356379098\\
912.168650929546	-12.2062286349671\\
922.361440293215	-12.1442267705245\\
932.731367333259	-12.081221614688\\
943.283554415979	-12.0172041472656\\
954.023321748267	-11.95216456751\\
964.956197042328	-11.8860925143181\\
976.08792575237	-11.8189769615305\\
987.424481923188	-11.7508062448157\\
998.97207969371	-11.6815679550424\\
1010.73718550214	-11.6112490135805\\
1022.72653104313	-11.5398357584017\\
1034.94712703154	-11.4673136068904\\
1047.40627783214	-11.3936673518494\\
1060.11159701926	-11.3188810211745\\
1073.07102393636	-11.242937951098\\
1086.292841331	-11.165820526302\\
1099.78569414793	-11.0875104332917\\
1113.55860956968	-11.0079883675103\\
1127.62101840256	-10.9272343765957\\
1141.98277791459	-10.8452274250477\\
1156.65419624165	-10.7619456097129\\
1171.64605848897	-10.6773660645668\\
1186.96965466693	-10.5914649567857\\
1202.63680961344	-10.5042173414849\\
1218.65991506951	-10.4155972886644\\
1235.05196409125	-10.3255776822895\\
1251.82658799898	-10.234130198888\\
1268.99809608466	-10.1412254705377\\
1286.58151832076	-10.0468326612837\\
1304.59265133869	-9.9509196750758\\
1323.04810797254	-9.85345309072854\\
1341.96537069498	-9.75439796856261\\
1361.36284930669	-9.65371782690913\\
1381.25994327971	-9.55137466708504\\
1401.67710919868	-9.44732872825858\\
1422.63593379306	-9.34153850309005\\
1444.15921310855	-9.23396069347876\\
1466.27103842855	-9.12454993523902\\
1488.99688962651	-9.01325882479907\\
1512.36373671018	-8.90003797966122\\
1536.40015040857	-8.78483536206976\\
1561.13642275559	-8.66759673941607\\
1586.60469874051	-8.54826529802855\\
1612.83912022874	-8.42678132506799\\
1639.87598350825	-8.30308252309491\\
1667.75391199063	-8.17710310375443\\
1696.51404579547	-8.04877436227659\\
1726.20025017518	-7.91802387951702\\
1756.85934500188	-7.78477576182539\\
1788.54135784156	-7.64894983979694\\
1821.29980349361	-7.51046195117772\\
1855.19199328173	-7.36922333032733\\
1890.27937785707	-7.22514037893311\\
1926.62792782796	-7.07811426546701\\
1964.308557177	-6.9280407128322\\
2003.39759518409	-6.77480925070279\\
2043.97731346456	-6.61830307688354\\
2086.13651578099	-6.45839844423421\\
2129.97119952842	-6.29496400984111\\
2175.58529926407	-6.12786035170991\\
2223.09152440442	-5.95693925288182\\
2272.61230530362	-5.78204293322358\\
2324.28086443301	-5.60300323139926\\
2378.24243239619	-5.41964065441294\\
2434.65563215366	-5.23176340995117\\
2493.69405924378	-5.03916620640275\\
2555.54809115959	-4.84162901712432\\
2620.42696561133	-4.63891555046527\\
2688.5611754765	-4.43077171834403\\
2760.20523820467	-4.2169237554406\\
2835.64090980885	-3.99707615114576\\
};
\addlegendentry{Coverage Radius = 500\,m}

\addplot [color=mycolor2,line width=1.0pt]
  table[row sep=crcr]{%
0	16.2815664386573\\
3.50235840561556	16.2347761881092\\
7.00486956737944	16.1871879619912\\
10.5076862680911	16.1387588240637\\
14.0109613438621	16.0894443882205\\
17.5148477107971	16.0391987101643\\
21.0194983917039	15.9879741566479\\
24.5250665428441	15.9357217452865\\
28.0317054807319	15.8823906582119\\
31.539568708994	15.8279284615122\\
35.0488099452976	15.7722810284751\\
38.5595831483589	15.7153925509319\\
42.0720425450409	15.6572053318658\\
45.586342657551	15.5976601404802\\
49.1026383307491	15.5366958077585\\
52.6210847595756	15.4742493878999\\
56.1418375166105	15.4102562311161\\
59.6650525797735	15.3446497079062\\
63.1908863601761	15.2773616003798\\
66.7194957301358	15.208321674354\\
70.2510380513641	15.1374580740794\\
73.7856712033377	15.0646970768249\\
77.3235536118655	14.9899632799263\\
80.8648442778614	14.9131793691822\\
84.4097028063345	14.8342666397362\\
87.958289435608	14.7531445466515\\
91.5107650667791	14.6697310479104\\
95.0672912934299	14.5839425363194\\
98.6280304316027	14.4956940290482\\
102.193145550052	14.4048991663747\\
105.762800500781	14.3114703462465\\
109.337159949888	14.2153188450625\\
112.916389408711	14.116354907705\\
116.500655265313	14.0144879017506\\
120.090124816298	13.909626576587\\
123.684966298981	13.8016789578364\\
127.285348923924	13.6905529371659\\
130.891442907852	13.576156008979\\
134.503419506965	13.4583959887605\\
138.121451050648	13.3371806801025\\
141.745710975617	13.2124187330633\\
145.376373860493	13.0840194596046\\
149.013615460829	12.9518933102347\\
152.657612744611	12.8159521578387\\
156.308543928233	12.6761096206087\\
159.96658851298	12.5322814034367\\
163.631927322024	12.3843856631607\\
167.304742537951	12.2323433864859\\
170.985217740842	12.0760787696969\\
174.673537946917	11.9155196628799\\
178.369889647771	11.7505979786826\\
182.074460850209	11.5812500750957\\
185.787441116705	11.4074172798058\\
189.509021606506	11.2290464129563\\
193.239395117393	11.0460900019078\\
196.97875612813	10.8585070139945\\
200.727300841612	10.6662631674525\\
204.485227228742	10.4693316165795\\
208.252735073045	10.2676930925436\\
212.030026016062	10.0613366189167\\
215.817303603527	9.85025996875983\\
219.614773332361	9.63446981194144\\
223.422642698511	9.41398261711255\\
227.241121245642	9.188824660124\\
231.070420614727	8.95903244608859\\
234.91075459455	8.72465329396958\\
238.76233917315	8.48574542251741\\
242.625392590236	8.24237818480833\\
246.500135390602	7.99463244657822\\
250.386790478566	7.74260084367198\\
254.285583173468	7.48638768227082\\
258.196741266257	7.22610891808267\\
262.120495077196	6.96189276066673\\
266.05707751472	6.69387891018324\\
270.006724135484	6.42221887631288\\
273.969673205619	6.14707574980913\\
277.946165763262	5.86862392374834\\
281.936445682362	5.58704877010317\\
285.940759737824	5.30254619346032\\
289.959357672022	5.01532234404051\\
293.992492262718	4.72559314311715\\
298.040419392429	4.4335835604254\\
302.103398119293	4.13952684153026\\
306.181690749462	3.84366435493261\\
310.275562911082	3.54624428521988\\
314.385283629895	3.24752110969814\\
318.511125406521	2.94775472204124\\
322.65336429546	2.64720952657282\\
326.812279985869	2.34615358744617\\
330.988155884168	2.04485750232089\\
335.181279198525	1.74359375197427\\
339.391941025282	1.44263556090586\\
343.620436437369	1.14225570349674\\
347.867064574781	0.842726044067334\\
352.132128737158	0.544316065222375\\
356.41593647856	0.247292328947879\\
360.718799704471	-0.0480827972982496\\
365.041034771121	-0.341551811708001\\
369.382962587195	-0.632862920593624\\
373.744908717985	-0.921770973035976\\
378.127203492078	-1.20803815017369\\
382.530182110645	-1.49143464940093\\
386.954184759421	-1.77173950325843\\
391.399556723442	-2.04874084159972\\
395.866648504644	-2.32223672724649\\
400.355815942398	-2.59203556934202\\
404.867420337078	-2.85795632518204\\
409.401828576749	-3.11982904594237\\
413.959413267088	-3.37749507893849\\
418.540552864621	-3.63080716342152\\
423.145631813394	-3.87962978143446\\
427.77504068517	-4.12383893056651\\
432.42917632329	-4.36332241195737\\
437.108441990287	-4.59797956408144\\
441.813247519386	-4.82772139037456\\
446.544009470014	-5.05247019668204\\
451.30115128745	-5.27215932823059\\
456.08510346674	-5.48673328330398\\
460.896303721026	-5.69614706681917\\
465.735197154428	-5.90036599672933\\
470.602236439621	-6.09936534109414\\
475.497882000277	-6.29312995509615\\
480.422602198514	-6.48165393802128\\
485.376873527535	-6.66494003970382\\
490.361180809609	-6.84299936286631\\
495.376017399599	-7.0158508255234\\
500.421885394192	-7.18352077404087\\
505.499295847053	-7.34604255978763\\
510.608768990077	-7.50345587873937\\
515.750834460971	-7.65580652271017\\
520.926031537352	-7.80314580677069\\
526.134909377623	-7.94553017321468\\
531.378027268815	-8.08302066868214\\
536.655954881684	-8.21568262584026\\
541.969272533268	-8.34358511531877\\
547.318571457208	-8.46680062284585\\
552.704454082068	-8.58540466235587\\
558.127534317962	-8.69947532118712\\
563.588437851774	-8.80909296721893\\
569.087802451267	-8.91433987052199\\
574.626278278419	-9.01529989447093\\
580.2045282123	-9.11205817294265\\
585.823228181841	-9.20470080375104\\
591.483067508854	-9.29331465863297\\
597.184749261685	-9.37798693429832\\
602.928990619873	-9.45880519373017\\
608.716523250235	-9.53585683293699\\
614.548093694791	-9.60922911551647\\
620.424463770982	-9.67900886873318\\
626.346410984629	-9.74528233912349\\
632.314728956114	-9.80813502041998\\
638.330227860294	-9.86765150930312\\
644.393734880656	-9.92391538368392\\
650.506094678268	-9.97700903639793\\
656.668169876094	-10.0270137648898\\
662.880841559253	-10.0740092789388\\
669.145009791866	-10.1180740351719\\
675.461594151121	-10.1592849785911\\
681.831534279233	-10.197717394862\\
688.255790454015	-10.233445024804\\
694.735344178798	-10.2665399598768\\
701.271198792466	-10.2970725391693\\
707.864380100417	-10.3251114071718\\
714.515937027304	-10.3507234110838\\
721.22694229243	-10.3739737723305\\
727.998493108729	-10.3949257331188\\
734.831711906301	-10.4136408671376\\
741.727747081519	-10.4301789485587\\
748.68777377277	-10.4445979295476\\
755.712994663946	-10.4569541077047\\
762.804640816846	-10.4673018830281\\
769.963972533732	-10.475693934774\\
777.192280251298	-10.4821812632141\\
784.490885467425	-10.4868131942269\\
791.861141702109	-10.489637182037\\
799.304435494082	-10.4906992687889\\
806.822187434636	-10.4900436266955\\
814.415853240329	-10.4877130077051\\
822.086924866273	-10.4837484401048\\
829.836931661801	-10.4781894767535\\
837.667441570431	-10.4710740957496\\
845.58006237611	-10.4624388764082\\
853.57644299784	-10.4523188969687\\
861.658274834896	-10.4407476821738\\
869.827293164959	-10.4277575685467\\
878.085278597616	-10.4133793615116\\
886.434058585799	-10.3976426098503\\
894.875508997881	-10.3805755420819\\
903.4115557533	-10.362205136064\\
912.044176524703	-10.342557171694\\
920.775402509836	-10.3216560429433\\
929.607320276488	-10.2995251604171\\
938.542073684072	-10.2761866672646\\
947.581865885563	-10.2516616797478\\
956.728961413755	-10.2259701513781\\
965.98568835601	-10.1991309672896\\
975.354440621901	-10.1711619903614\\
984.837680308442	-10.1420800882836\\
994.437940167809	-10.1119011521996\\
1004.1578261828	-10.0806400816272\\
1014.00002025558	-10.0483108276485\\
1023.96728301547	-10.0149264277001\\
1034.06245675215	-9.98049901443852\\
1044.28846848073	-9.94503999330479\\
1054.64833314568	-9.90855967607969\\
1065.14515697104	-9.87106775388985\\
1075.78214096481	-9.83257294053738\\
1086.56258458573	-9.79308325804934\\
1097.48988958151	-9.75260591675244\\
1108.56756400767	-9.7111473639041\\
1119.79922643728	-9.66871329958182\\
1131.18861037197	-9.62530868954448\\
1142.73956886585	-9.58093777833743\\
1154.45607937408	-9.53560406458224\\
1166.34224883927	-9.48931046102513\\
1178.40231902917	-9.44205912306161\\
1190.64067214044	-9.39385145528652\\
1203.06183668392	-9.34468827173377\\
1215.67049366832	-9.29456980580323\\
1228.47148309981	-9.24349547474725\\
1241.46981081673	-9.1914641512339\\
1254.67065567974	-9.13847404933324\\
1268.079377139	-9.084522742747\\
1281.70152320185	-9.02960716937814\\
1295.54283882585	-8.97372362101979\\
1309.60927476394	-8.91686783529964\\
1323.90699689031	-8.85903480840548\\
1338.44239603786	-8.80021895336563\\
1353.22209838032	-8.74041410409548\\
1368.25297639432	-8.67961348116216\\
1383.54216043982	-8.61780949426286\\
1399.09705099989	-8.55499406198491\\
1414.92533162397	-8.49115849848089\\
1431.0349826224	-8.42629332377265\\
1447.43429556349	-8.36038848449188\\
1464.13188862856	-8.29343324969857\\
1481.13672288478	-8.22541623826505\\
1498.45811954056	-8.15632531272924\\
1516.10577825321	-8.08614765505746\\
1534.08979656469	-8.01486985318908\\
1552.42069054731	-7.94247756427475\\
1571.10941674821	-7.86895581109165\\
1590.1673955289	-7.79428884211336\\
1609.60653590454	-7.71846020513144\\
1629.4392619965	-7.64145248773315\\
1649.6785412219	-7.56324757102431\\
1670.33791435452	-7.4838263370802\\
1691.43152760384	-7.40316901252488\\
1712.97416687188	-7.32125473350834\\
1734.98129436247	-7.23806176148871\\
1757.46908773345	-7.15356738830062\\
1780.4544820004	-7.06774793250113\\
1803.95521442016	-6.98057859437151\\
1827.98987260427	-6.89203358311589\\
1852.57794613688	-6.80208591618478\\
1877.73988199847	-6.71070739810521\\
1903.49714412699	-6.61786878369199\\
1929.87227748114	-6.52353935468004\\
1956.88897700804	-6.42768712786742\\
1984.57216195881	-6.3302787902737\\
2012.94805604248	-6.231279505971\\
2042.04427396004	-6.1306528927704\\
2071.88991491956	-6.02836104737221\\
2102.51566379802	-5.92436430039926\\
2133.95390068958	-5.81862123219857\\
2166.23881966282	-5.7110886287427\\
2199.40655764282	-5.60172120645396\\
2233.49533443977	-5.49047163904552\\
2268.54560506526	-5.37729061811784\\
2304.60022561285	-5.26212617695749\\
2341.70463413339	-5.14492415106773\\
2379.90704811077	-5.02562779207886\\
2419.25868034312	-4.90417744975899\\
2459.81397526237	-4.78051088669234\\
2501.63086798595	-4.65456237207013\\
2544.7710686932	-4.52626325629302\\
2589.30037526278	-4.39554117310939\\
2635.28901750282	-4.26232027957764\\
2682.81203676234	-4.12652045490759\\
2731.94970524042	-3.98805758345195\\
2782.78798992259	-3.84684294425176\\
2835.4190667856	-3.70278298183001\\
2889.94189174194	-3.55577890572442\\
2946.4628357655	-3.40572647820704\\
3005.09639277614	-3.25251526669403\\
3065.96597019683	-3.09602850517231\\
3129.20477367149	-2.93614248118837\\
3194.95679929263	-2.77272588508137\\
3263.3779488961	-2.60563932673526\\
3334.63728660663	-2.43473461975151\\
3408.91845795543	-2.25985401329365\\
3486.42129664951	-2.08082937411112\\
3567.36364859428	-1.89748123613202\\
3651.98344823048	-1.70961783285345\\
3740.54108886567	-1.51703389740736\\
3833.32213673939	-1.31950942687011\\
3930.640448417	-1.11680815232977\\
4032.84176321476	-0.908676007502834\\
4140.30785730701	-0.694839247965675\\
4253.46136471328	-0.475002383140975\\
};
\addlegendentry{Coverage Radius = 750\,m}

\addplot [color=mycolor3,line width=1.0pt]
  table[row sep=crcr]{%
0	20.0019209382852\\
4.66981120748742	19.953957035731\\
9.33982608983926	19.9051468470078\\
14.0102483574548	19.8554455525808\\
18.6812817918162	19.8048068186705\\
23.3531302810628	19.7531826875151\\
28.0259978556053	19.700523445241\\
32.7000887237921	19.6467779603565\\
37.3756073076426	19.5918931961496\\
42.0527582786587	19.5358144289431\\
46.7317465937302	19.4784851702367\\
51.4127775311453	19.4198471770466\\
56.0960567267212	19.3598402436179\\
60.781790210068	19.2984025557208\\
65.4701844409988	19.2354702855758\\
70.1614463461008	19.1709777528115\\
74.8557833554806	19.1048574969535\\
79.553403439698	19.0370400015829\\
84.2545151469015	18.9674540857517\\
88.9593276401811	18.8960264764029\\
93.6680507351522	18.8226822036709\\
98.3808949377836	18.7473443563153\\
103.098071482487	18.6699342699154\\
107.819792370482	18.5903712967357\\
112.546270408446	18.5085733284244\\
117.277719247477	18.42445634878\\
122.014353422372	18.3379347793438\\
126.75638839124	18.2489214143281\\
131.50404057547	18.1573276136356\\
136.257527400069	18.063063305573\\
141.017067334375	17.9660371258966\\
145.782879933184	17.8661565436227\\
150.555185878281	17.7633279566639\\
155.334207020417	17.6574568522456\\
160.120166421731	17.5484480728588\\
164.913288398641	17.436205718177\\
169.713798565232	17.3206337430397\\
174.521923877136	17.2016357030244\\
179.337892675953	17.0791154823953\\
184.161934734197	16.9529769714776\\
188.994281300823	16.8231249361226\\
193.835165147324	16.6894648450197\\
198.684820614439	16.5519033589911\\
203.543483659482	16.4103486282477\\
208.411391904311	16.2647106295207\\
213.288784683974	16.114901522415\\
218.175903096032	15.9608360303635\\
223.072990050602	15.8024318349747\\
227.980290321122	15.6396099728592\\
232.898050595889	15.4722952976131\\
237.826519530361	15.3004169069335\\
242.765947800278	15.1239085442908\\
247.716588155607	14.9427091426372\\
252.678695475341	14.7567633669067\\
257.65252682319	14.5660218488075\\
262.638341504173	14.3704419397977\\
267.63640112215	14.1699880423124\\
272.646969638323	13.9646323151709\\
277.670313430727	13.7543548344531\\
282.70670135475	13.5391443287724\\
287.756404804702	13.3189986556207\\
292.819697776482	13.0939249673033\\
297.896856931349	12.8639406307192\\
302.988161660857	12.6290732525967\\
308.09389415297	12.3893611172066\\
313.2143394594	12.1448537852528\\
318.3497855642	11.8956121925797\\
323.500523453648	11.6417088957664\\
328.666847187469	11.383228459625\\
333.849053971421	11.1202677217503\\
339.047444231291	10.8529356976201\\
344.262321688342	10.5813535622798\\
349.493993436261	10.3056552546437\\
354.742770019627	10.0259867115834\\
360.008965513978	9.74250618144431\\
365.292897607492	9.45538398737362\\
370.594887684349	9.16480223746719\\
375.915260909815	8.87095448732958\\
381.254346317098	8.57404527686472\\
386.612476896029	8.2742898234799\\
391.989989683624	7.97191352485457\\
397.387225856572	7.6671512109511\\
402.804530825724	7.36024634335141\\
408.242254332617	7.05145083184449\\
413.700750548109	6.74102369474582\\
419.180378173193	6.42923050058826\\
424.681500542028	6.11634245494784\\
430.20448572728	5.80263545627196\\
435.749706647825	5.48838920524668\\
441.31754117889	5.17388603759159\\
446.9083722647	4.8594102320225\\
452.522588033709	4.54524682875429\\
458.160581916493	4.23168039342312\\
463.822752766374	3.91899451509283\\
469.509504982878	3.60747029305237\\
475.22124863808	3.29738575718017\\
480.958399605961	2.98901455441857\\
486.721379694828	2.68262523943872\\
492.510616782927	2.37848039562864\\
498.326544957313	2.07683566582332\\
504.169604656104	1.7779390305376\\
510.040242814194	1.48203009363277\\
515.938913012561	1.18933923580217\\
521.866075631256	0.90008732743479\\
527.822198006192	0.614484870753884\\
533.807754589865	0.332731567102158\\
539.823227116104	0.0550160986441739\\
545.869104768998	-0.21848443091281\\
551.945884356117	-0.48760470922312\\
558.054070486162	-0.752191318871795\\
564.194175751191	-1.01210309319353\\
570.36672091356	-1.26721089126362\\
576.57223509772	-1.51739788551799\\
582.811255987049	-1.76255929265734\\
589.084330025848	-2.00260249603743\\
595.392012626685	-2.23744667567898\\
601.734868383266	-2.46702253576417\\
608.113471288986	-2.69127240697276\\
614.528404961368	-2.91014958703337\\
620.98026287257	-3.12361813259848\\
627.469648586161	-3.33165248046643\\
633.997176000369	-3.53423706761166\\
640.563469598019	-3.73136597021995\\
647.16916470338	-3.92304229140376\\
653.814907746146	-4.10927784419592\\
660.501356532798	-4.29009259415203\\
667.229180525589	-4.46551425230437\\
673.999061129404	-4.63557783154873\\
680.81169198677	-4.80032496595494\\
687.667779281294	-4.95980364180058\\
694.568042049803	-5.1140676049975\\
701.513212503497	-5.26317594537416\\
708.50403635842	-5.4071925545116\\
715.541273175578	-5.54618578861377\\
722.625696711024	-5.68022790217742\\
729.758095276277	-5.80939470746685\\
736.939272109424	-5.93376517061319\\
744.170045757283	-6.05342094049716\\
751.451250469031	-6.16844604029595\\
758.783736601689	-6.27892647429107\\
766.168371037893	-6.38494990490302\\
773.606037616401	-6.48660531658527\\
781.097637575787	-6.58398269673138\\
788.644090011805	-6.67717283390832\\
796.246332348913	-6.76626685792102\\
803.905320826498	-6.8513562708833\\
811.622031000313	-6.93253240448166\\
819.397458259721	-7.00988644577296\\
827.23261836131	-7.08350912519727\\
835.128547979505	-7.15349056479565\\
843.086305274819	-7.21992009939999\\
851.106970480392	-7.28288612627597\\
859.191646507541	-7.34247597789511\\
867.341459571025	-7.39877575068752\\
875.557559834792	-7.45187039032618\\
883.841122079003	-7.50184319556318\\
892.193346389155	-7.54877614933679\\
900.615458868162	-7.59274965734546\\
909.10871237231	-7.63384239780845\\
917.674387272019	-7.67213143382612\\
926.313792238398	-7.70769210710617\\
935.028265056621	-7.74059793353618\\
943.819173467222	-7.77092065985801\\
952.687916036406	-7.79873016017204\\
961.635923056574	-7.82409460692087\\
970.664657478306	-7.84708011715952\\
979.775615875068	-7.86775106319814\\
988.970329442025	-7.88616994175088\\
998.25036503036	-7.90239735178227\\
1007.61732621859	-7.91649216245305\\
1017.07285442246	-7.92851127073972\\
1026.61863004498	-7.93850977908824\\
1036.2563736684	-7.94654103808439\\
1045.9878472899	-7.95265665206715\\
1055.81485560281	-7.95690628298173\\
1065.73924732544	-7.9593381111512\\
1075.76291657951	-7.95999837869118\\
1085.88780432044	-7.95893184050487\\
1096.1158998217	-7.95618146226398\\
1106.44924221573	-7.95178867006116\\
1116.88992209391	-7.94579325253062\\
1127.44008316815	-7.93823353830463\\
1138.10192399712	-7.92914629521854\\
1148.87769977986	-7.91856667940223\\
1159.76972421994	-7.90652860207474\\
1170.78037146349	-7.89306438818677\\
1181.9120781144	-7.87820505239553\\
1193.16734533051	-7.86198023695701\\
1204.5487410044	-7.84441828282971\\
1216.05890203294	-7.82554628386448\\
1227.70053667978	-7.80538990031218\\
1239.47642703532	-7.78397376283669\\
1251.38943157876	-7.76132118985859\\
1263.44248784742	-7.73745442953343\\
1275.63861521834	-7.71239452527367\\
1287.98091780801	-7.68616141148091\\
1300.47258749587	-7.65877396099995\\
1313.11690707792	-7.63025001348824\\
1325.91725355708	-7.60060639533221\\
1338.87710157707	-7.56985890581279\\
1352.00002700744	-7.53802236150963\\
1365.28971068729	-7.50511063227475\\
1378.7499423362	-7.47113665125067\\
1392.38462464098	-7.43611259355689\\
1406.19777752758	-7.40004951093614\\
1420.19354262806	-7.36295780582054\\
1434.37618795307	-7.32484687569074\\
1448.75011278097	-7.28572539962455\\
1463.31985277534	-7.24560121934088\\
1478.09008534356	-7.20448138876995\\
1493.0656352497	-7.16237219085318\\
1508.25148049596	-7.11927915128572\\
1523.6527584878	-7.07520705247474\\
1539.27477249877	-7.03015990965451\\
1555.12299845236	-6.98414113173314\\
1571.2030920389	-6.93715335058884\\
1587.52089618725	-6.88919842836546\\
1604.0824489119	-6.84027761843261\\
1620.89399155776	-6.79039157600788\\
1637.96197746641	-6.73954012331477\\
1655.29308108898	-6.68772252179376\\
1672.89420757299	-6.63493735871349\\
1690.772502852	-6.58118256600316\\
1708.93536426913	-6.5264554254033\\
1727.39045176781	-6.47075255871482\\
1746.14569968526	-6.41407002028313\\
1765.20932918707	-6.35640311024393\\
1784.58986138382	-6.29774653330452\\
1804.29613117376	-6.2380944032728\\
1824.33730185909	-6.17744020928643\\
1844.72288058643	-6.11577661873841\\
1865.46273466652	-6.05309579746748\\
1886.56710883196	-5.98938929684743\\
1908.04664349653	-5.92464786448964\\
1929.91239408466	-5.85886166536689\\
1952.17585150474	-5.79202017800067\\
1974.84896384638	-5.72411222219976\\
1997.94415938742	-5.65512585325439\\
2021.47437100429	-5.585048438027\\
2045.45306208626	-5.51386674181185\\
2069.89425406308	-5.44156659187634\\
2094.81255566428	-5.36813317416724\\
2120.22319403853	-5.29355089366044\\
2146.14204787271	-5.21780344825122\\
2172.585682662	-5.14087356949032\\
2199.57138829587	-5.06274327655495\\
2227.11721913936	-4.98339358393797\\
2255.24203680511	-4.90280484525607\\
2283.96555582917	-4.82095631844696\\
2313.3083924833	-4.73782638176262\\
2343.29211697793	-4.65339243904043\\
2373.93930933386	-4.56763091624386\\
2405.27361922687	-4.48051711665104\\
2437.31983013903	-4.39202534823305\\
2470.10392818251	-4.30212872314962\\
2503.65317599796	-4.21079913674465\\
2537.99619216932	-4.11800743091634\\
2573.16303664153	-4.02372297090173\\
2609.18530267738	-3.92791385356611\\
2646.09621594508	-3.83054684270175\\
2683.93074138997	-3.7315871759935\\
2722.72569861338	-3.6309985418346\\
2762.51988655942	-3.52874310460101\\
2803.35421839737	-3.42478125980357\\
2845.27186758611	-3.31907165000401\\
2888.3184262171	-3.21157112082585\\
2932.54207685709	-3.10223444588368\\
2977.99377925302	-2.99101435372516\\
3024.72747342035	-2.87786158852427\\
3072.80030081714	-2.76272423397286\\
3122.27284551118	-2.64554817390034\\
3173.20939748102	-2.52627670626941\\
3225.67824045749	-2.40485022526887\\
3279.75196701649	-2.28120653607033\\
3335.50782398126	-2.1552799486947\\
3393.02809159093	-2.02700185268677\\
3452.40050035037	-1.89629991932294\\
3513.71869000377	-1.76309834164003\\
3577.08271568312	-1.62731703334043\\
3642.59960698723	-1.48887191184483\\
3710.38398656346	-1.34767428789627\\
3780.55875571413	-1.20363063641005\\
3853.25585565592	-1.05664219606057\\
3928.61711435401	-0.906604757052534\\
4006.79519036819	-0.753407913580503\\
4087.95462692911	-0.596934925303088\\
4172.27303156198	-0.437062104377347\\
4259.94239905684	-0.273658164735764\\
4351.17059852814	-0.106583738880232\\
4446.18304880884	0.0643093379048594\\
4545.22461060725	0.239178795669631\\
4648.56172886601	0.41819274773772\\
4756.48486479237	0.60153064107817\\
4869.31126430731	0.789384223880973\\
4987.38811848756	0.981958745458655\\
5111.09618231918	1.17947419190541\\
5240.85393122266	1.38216681599987\\
5377.12235095301	1.59029066855952\\
5520.41047640934	1.80411947918181\\
5671.28181961771	2.0239487242312\\
};
\addlegendentry{Coverage Radius = 1000\,m}

\end{axis}
\end{tikzpicture}%

%% file: CovRadius_Power.tex
%
%
\pgfplotsset{every tick label/.append style={font=\small}}
\pgfplotsset{every axis/.append style={
        scaled y ticks = false, 
        y tick label style={/pgf/number format/.cd, fixed, fixed zerofill,
                          int detect,1000 sep={},precision=3},
    }
}
\definecolor{mycolor1}{rgb}{0.00000,0.44700,0.74100}%
\definecolor{mycolor2}{rgb}{0.85000,0.32500,0.09800}%
\definecolor{mycolor3}{rgb}{0.92900,0.69400,0.12500}%
\begin{tikzpicture}

\begin{axis}[%
legend style={font=\fontsize{9}{5}\selectfont}, 
width=7cm,
height=7cm,
at={(0.974in,0.632in)},
scale only axis,
xmin=0,
xmax=4000,
xlabel style={font=\color{white!15!black}},
xlabel={Altitude, $h$ [m]},
ymin=0,
ymax=2500,
ylabel style={font=\color{white!15!black}},
ylabel={Coverage Radius [m]},
axis background/.style={fill=white},
xmajorgrids,
ymajorgrids,
legend style={at={(0.404,0.568)}, anchor=south west, legend cell align=left, align=left, draw=white!15!black}
]
\addplot [color=mycolor1,line width=1.0pt]
  table[row sep=crcr]{%
0	94.990234375\\
10	102.268721938133\\
20	112.281728037867\\
30	125.070958832707\\
40	139.603854624439\\
50	154.940659622186\\
60	170.540169209574\\
70	186.128931830945\\
80	201.57248406779\\
90	216.805684179644\\
100	231.798163126723\\
110	246.537379822611\\
120	261.020453993793\\
130	275.249008084218\\
140	289.227546767633\\
150	302.961563341677\\
160	316.457154359164\\
170	329.720350489424\\
180	342.757085927848\\
190	355.572950983463\\
200	368.173325994176\\
210	380.562996604948\\
220	392.746517466411\\
230	404.728069921582\\
240	416.511461259365\\
250	428.100166875315\\
260	439.497380104009\\
270	450.705951479609\\
280	461.728427122522\\
290	472.567184200931\\
300	483.22439046756\\
310	493.701834593607\\
320	504.001257460176\\
330	514.12409918349\\
340	524.07174074566\\
350	533.84530766368\\
360	543.445680899968\\
370	552.873689375229\\
380	562.130027546069\\
390	571.21523453424\\
400	580.129631983201\\
410	588.873546366783\\
420	597.447114706711\\
430	605.850310113186\\
440	614.083152624175\\
450	622.145301405571\\
460	630.03656987178\\
470	637.756467009121\\
480	645.304559929923\\
490	652.680191853888\\
500	659.882744241143\\
510	666.911291926129\\
520	673.76512938193\\
530	680.443113250676\\
540	686.944329500239\\
550	693.267558973674\\
560	699.411639541551\\
570	705.375184049197\\
580	711.156890508114\\
590	716.755191969047\\
600	722.168642609075\\
610	727.395566261931\\
620	732.434314464351\\
630	737.283048401887\\
640	741.939934843948\\
650	746.403138961732\\
660	750.670532854092\\
670	754.740253928358\\
680	758.609987764718\\
690	762.277747712467\\
700	765.741080287194\\
710	768.997854708653\\
720	772.045608314113\\
730	774.881980019081\\
740	777.504490138225\\
750	779.910686561706\\
760	782.097848355322\\
770	784.063505274378\\
780	785.804919818897\\
790	787.319536139932\\
800	788.604609788483\\
810	789.657360121108\\
820	790.475049980135\\
830	791.054915825286\\
840	791.394248883775\\
850	791.490175367914\\
860	791.339889938994\\
870	790.940663939373\\
880	790.289703214562\\
890	789.384157905102\\
900	788.22128396109\\
910	786.798378927588\\
920	785.112643951679\\
930	783.16134804046\\
940	780.941916867291\\
950	778.451499695991\\
960	775.687510458665\\
970	772.647272526\\
980	769.328188001496\\
990	765.727538153096\\
1000	761.842802297313\\
1010	757.671243312871\\
1020	753.210158480595\\
1030	748.456838185534\\
1040	743.408600396316\\
1050	738.062472664031\\
1060	732.415595381703\\
1070	726.464774548102\\
1080	720.206961021848\\
1090	713.638613864419\\
1100	706.756124459333\\
1110	699.55562046382\\
1120	692.032990473027\\
1130	684.183588644079\\
1140	676.002559637944\\
1150	667.484513240041\\
1160	658.62343123816\\
1170	649.412918818469\\
1180	639.845595297221\\
1190	629.913665854533\\
1200	619.608160153117\\
1210	608.919231577768\\
1220	597.836020741166\\
1230	586.346200078106\\
1240	574.436098807351\\
1250	562.090370429901\\
1260	549.29188120337\\
1270	536.021196880153\\
1280	522.256444333583\\
1290	507.97290669598\\
1300	493.142529325292\\
1310	477.733142116497\\
1320	461.707894548198\\
1330	445.024473318173\\
1340	427.63365964957\\
1350	409.478021609272\\
1360	390.489771721865\\
1370	370.588265308841\\
1380	349.676515065637\\
1390	327.635985900147\\
1400	304.319993133717\\
1410	279.542624369031\\
1420	253.063578860809\\
1430	224.562165863966\\
1440	193.594664215257\\
1450	159.515315411764\\
1460	121.319300602903\\
1470	77.2776510269645\\
1480	23.8884939652651\\
1490	0.000145803796114318\\
1500	0\\
1510	0\\
1520	0\\
1530	0\\
1540	0\\
1550	0\\
1560	0\\
1570	0\\
1580	0\\
1590	0\\
1600	0\\
1610	0\\
1620	0\\
1630	0\\
1640	0\\
1650	0\\
1660	0\\
1670	0\\
1680	0\\
1690	0\\
1700	0\\
1710	0\\
1720	0\\
1730	0\\
1740	0\\
1750	0\\
1760	0\\
1770	0\\
1780	0\\
1790	0\\
1800	0\\
1810	0\\
1820	0\\
1830	0\\
1840	0\\
1850	0\\
1860	0\\
1870	0\\
1880	0\\
1890	0\\
1900	0\\
1910	0\\
1920	0\\
1930	0\\
1940	0\\
1950	0\\
1960	0\\
1970	0\\
1980	0\\
1990	0\\
2000	0\\
2010	0\\
2020	0\\
2030	0\\
2040	0\\
2050	0\\
2060	0\\
2070	0\\
2080	0\\
2090	0\\
2100	0\\
2110	0\\
2120	0\\
2130	0\\
2140	0\\
2150	0\\
2160	0\\
2170	0\\
2180	0\\
2190	0\\
2200	0\\
2210	0\\
2220	0\\
2230	0\\
2240	0\\
2250	0\\
2260	0\\
2270	0\\
2280	0\\
2290	0\\
2300	0\\
2310	0\\
2320	0\\
2330	0\\
2340	0\\
2350	0\\
2360	0\\
2370	0\\
2380	0\\
2390	0\\
2400	0\\
2410	0\\
2420	0\\
2430	0\\
2440	0\\
2450	0\\
2460	0\\
2470	0\\
2480	0\\
2490	0\\
2500	0\\
2510	0\\
2520	0\\
2530	0\\
2540	0\\
2550	0\\
2560	0\\
2570	0\\
2580	0\\
2590	0\\
2600	0\\
2610	0\\
2620	0\\
2630	0\\
2640	0\\
2650	0\\
2660	0\\
2670	0\\
2680	0\\
2690	0\\
2700	0\\
2710	0\\
2720	0\\
2730	0\\
2740	0\\
2750	0\\
2760	0\\
2770	0\\
2780	0\\
2790	0\\
2800	0\\
2810	0\\
2820	0\\
2830	0\\
2840	0\\
2850	0\\
2860	0\\
2870	0\\
2880	0\\
2890	0\\
2900	0\\
2910	0\\
2920	0\\
2930	0\\
2940	0\\
2950	0\\
2960	0\\
2970	0\\
2980	0\\
2990	0\\
3000	0\\
3010	0\\
3020	0\\
3030	0\\
3040	0\\
3050	0\\
3060	0\\
3070	0\\
3080	0\\
3090	0\\
3100	0\\
3110	0\\
3120	0\\
3130	0\\
3140	0\\
3150	0\\
3160	0\\
3170	0\\
3180	0\\
3190	0\\
3200	0\\
3210	0\\
3220	0\\
3230	0\\
3240	0\\
3250	0\\
3260	0\\
3270	0\\
3280	0\\
3290	0\\
3300	0\\
3310	0\\
3320	0\\
3330	0\\
3340	0\\
3350	0\\
3360	0\\
3370	0\\
3380	0\\
3390	0\\
3400	0\\
3410	0\\
3420	0\\
3430	0\\
3440	0\\
3450	0\\
3460	0\\
3470	0\\
3480	0\\
3490	0\\
3500	0\\
3510	0\\
3520	0\\
3530	0\\
3540	0\\
3550	0\\
3560	0\\
3570	0\\
3580	0\\
3590	0\\
3600	0\\
3610	0\\
3620	0\\
3630	0\\
3640	0\\
3650	0\\
3660	0\\
3670	0\\
3680	0\\
3690	0\\
3700	0\\
3710	0\\
3720	0\\
3730	0\\
3740	0\\
3750	0\\
3760	0\\
3770	0\\
3780	0\\
3790	0\\
3800	0\\
3810	0\\
3820	0\\
3830	0\\
3840	0\\
3850	0\\
3860	0\\
3870	0\\
3880	0\\
3890	0\\
3900	0\\
3910	0\\
3920	0\\
3930	0\\
3940	0\\
3950	0\\
3960	0\\
3970	0\\
3980	0\\
3990	0\\
4000	0\\
};
\addlegendentry{ADr's P$_\text{Tx}$ = -10\,dBm}

\addplot [color=mycolor2,line width=1.0pt]
  table[row sep=crcr]{%
0	139.426574707031\\
10	146.676443852135\\
20	155.735621125394\\
30	166.967734369906\\
40	180.140748752224\\
50	194.706817108361\\
60	210.135392307564\\
70	226.039354292048\\
80	242.167370522112\\
90	258.36334807641\\
100	274.531458794904\\
110	290.613202166488\\
120	306.573071183782\\
130	322.38975018931\\
140	338.050945537185\\
150	353.550062470309\\
160	368.884020572571\\
170	384.052304548385\\
180	399.055739485919\\
190	413.896188040482\\
200	428.576063215672\\
210	443.098200544479\\
220	457.465458215103\\
230	471.680979541626\\
240	485.747794520834\\
250	499.66877590239\\
260	513.446912142816\\
270	527.084954017703\\
280	540.585745396196\\
290	553.951758058813\\
300	567.185411641741\\
310	580.289176625096\\
320	593.265184533838\\
330	606.115616125598\\
340	618.842427873075\\
350	631.447484962879\\
360	643.932738831924\\
370	656.299711444704\\
380	668.550126991334\\
390	680.685488512187\\
400	692.707204277478\\
410	704.616676683173\\
420	716.41521070379\\
430	728.103922887019\\
440	739.684067704733\\
450	751.156698971825\\
460	762.522878323543\\
470	773.783474139728\\
480	784.939538348423\\
490	795.99174899032\\
500	806.941099146233\\
510	817.788264134556\\
520	828.533932794624\\
530	839.178854233422\\
540	849.723570749434\\
550	860.168700110754\\
560	870.514832099667\\
570	880.76241923974\\
580	890.911997922297\\
590	900.964002503957\\
600	910.91882575016\\
610	920.7768853003\\
620	930.538432548904\\
630	940.203882967478\\
640	949.773541229489\\
650	959.247581577393\\
660	968.626211458615\\
670	977.909753805489\\
680	987.09836163205\\
690	996.192187537887\\
700	1005.19137393662\\
710	1014.09594752342\\
720	1022.90609562252\\
730	1031.62187255365\\
740	1040.24328447932\\
750	1048.77037944184\\
760	1057.20319238212\\
770	1065.5416373611\\
780	1073.78575078031\\
790	1081.93538336816\\
800	1089.99054880288\\
810	1097.95111305553\\
820	1105.81699350095\\
830	1113.58799814998\\
840	1121.26402992243\\
850	1128.84492561043\\
860	1136.33044976795\\
870	1143.72045129011\\
880	1151.01475377081\\
890	1158.21298675471\\
900	1165.31501841631\\
910	1172.32057123016\\
920	1179.2293265039\\
930	1186.04097684605\\
940	1192.75522411149\\
950	1199.37177737024\\
960	1205.89023652061\\
970	1212.3102597191\\
980	1218.63150524161\\
990	1224.85351339475\\
1000	1230.97593534581\\
1010	1236.99829953145\\
1020	1242.92024297077\\
1030	1248.74127537585\\
1040	1254.46089378918\\
1050	1260.07870066732\\
1060	1265.59416469897\\
1070	1271.0067976713\\
1080	1276.31603312441\\
1090	1281.52140625088\\
1100	1286.62231061381\\
1110	1291.61823975953\\
1120	1296.50854201103\\
1130	1301.29266416925\\
1140	1305.97009061796\\
1150	1310.54009417851\\
1160	1315.00204415945\\
1170	1319.35540752653\\
1180	1323.59943541729\\
1190	1327.73353777224\\
1200	1331.75696889813\\
1210	1335.66907865789\\
1220	1339.4691225068\\
1230	1343.1564514134\\
1240	1346.73032079174\\
1250	1350.19001737228\\
1260	1353.53473065928\\
1270	1356.76374517872\\
1280	1359.87624745708\\
1290	1362.87151931255\\
1300	1365.74874393175\\
1310	1368.50713501633\\
1320	1371.14587188512\\
1330	1373.66409919538\\
1340	1376.06105773944\\
1350	1378.33588859627\\
1360	1380.48776386943\\
1370	1382.51575549947\\
1380	1384.41909845663\\
1390	1386.19686212476\\
1400	1387.84821387258\\
1410	1389.37222150375\\
1420	1390.76798556676\\
1430	1392.0347063048\\
1440	1393.17141906312\\
1450	1394.17719333626\\
1460	1395.0511334886\\
1470	1395.79237946317\\
1480	1396.39997436483\\
1490	1396.87299803779\\
1500	1397.21056786363\\
1510	1397.41163967406\\
1520	1397.47540820105\\
1530	1397.40090823603\\
1540	1397.18714871294\\
1550	1396.8333135619\\
1560	1396.3382962629\\
1570	1395.70136712557\\
1580	1394.92144157314\\
1590	1393.99761421575\\
1600	1392.92896017237\\
1610	1391.71460278775\\
1620	1390.35358199438\\
1630	1388.84492234015\\
1640	1387.18776699367\\
1650	1385.38118014197\\
1660	1383.42421507313\\
1670	1381.31591591724\\
1680	1379.05545120517\\
1690	1376.64185165297\\
1700	1374.07427575072\\
1710	1371.3516829973\\
1720	1368.47329562327\\
1730	1365.43807610305\\
1740	1362.24525305656\\
1750	1358.893865957\\
1760	1355.38296420362\\
1770	1351.71167431707\\
1780	1347.87907278994\\
1790	1343.88425392276\\
1800	1339.72633261731\\
1810	1335.40438335913\\
1820	1330.91738051559\\
1830	1326.26452089121\\
1840	1321.44471775197\\
1850	1316.45717685154\\
1860	1311.30076749837\\
1870	1305.97453354027\\
1880	1300.47757013227\\
1890	1294.80871744294\\
1900	1288.96700346086\\
1910	1282.951338146\\
1920	1276.76064425907\\
1930	1270.39368063711\\
1940	1263.84935573838\\
1950	1257.12642965916\\
1960	1250.22364532333\\
1970	1243.13967738365\\
1980	1235.87320198213\\
1990	1228.42278801911\\
2000	1220.78696868886\\
2010	1212.96413669699\\
2020	1204.95273354327\\
2030	1196.7510877899\\
2040	1188.35731769108\\
2050	1179.76957926245\\
2060	1170.98591157566\\
2070	1162.00425746539\\
2080	1152.8223741535\\
2090	1143.43791478916\\
2100	1133.84850829219\\
2110	1124.05145882009\\
2120	1114.04405240297\\
2130	1103.82331953427\\
2140	1093.38623431948\\
2150	1082.72948906235\\
2160	1071.84954336086\\
2170	1060.74267501437\\
2180	1049.40498214172\\
2190	1037.83219016313\\
2200	1026.01987321451\\
2210	1013.96322775268\\
2220	1001.65711085793\\
2230	989.096180750998\\
2240	976.274518176141\\
2250	963.185939428326\\
2260	949.823839871325\\
2270	936.180973927937\\
2280	922.249717937931\\
2290	908.021891807859\\
2300	893.488619438453\\
2310	878.64040057971\\
2320	863.46695751241\\
2330	847.957299883431\\
2340	832.099342940631\\
2350	815.880176337007\\
2360	799.285746095621\\
2370	782.300725803934\\
2380	764.908348703791\\
2390	747.090449140065\\
2400	728.826925612803\\
2410	710.095936836053\\
2420	690.873201216839\\
2430	671.131867931069\\
2440	650.842241066303\\
2450	629.971178046574\\
2460	608.481435711349\\
2470	586.331183191502\\
2480	563.473137703896\\
2490	539.853314218325\\
2500	515.409848279309\\
2510	490.071315184142\\
2520	463.75435993012\\
2530	436.360825224867\\
2540	407.774000998333\\
2550	377.853064126602\\
2560	346.425390933074\\
2570	313.276023719403\\
2580	278.130823928595\\
2590	240.631479950286\\
2600	200.293689442262\\
2610	156.434823785064\\
2620	108.036964724467\\
2630	53.4574933013245\\
2640	0.000163139322733719\\
2650	0\\
2660	0\\
2670	0\\
2680	0\\
2690	0\\
2700	0\\
2710	0\\
2720	0\\
2730	0\\
2740	0\\
2750	0\\
2760	0\\
2770	0\\
2780	0\\
2790	0\\
2800	0\\
2810	0\\
2820	0\\
2830	0\\
2840	0\\
2850	0\\
2860	0\\
2870	0\\
2880	0\\
2890	0\\
2900	0\\
2910	0\\
2920	0\\
2930	0\\
2940	0\\
2950	0\\
2960	0\\
2970	0\\
2980	0\\
2990	0\\
3000	0\\
3010	0\\
3020	0\\
3030	0\\
3040	0\\
3050	0\\
3060	0\\
3070	0\\
3080	0\\
3090	0\\
3100	0\\
3110	0\\
3120	0\\
3130	0\\
3140	0\\
3150	0\\
3160	0\\
3170	0\\
3180	0\\
3190	0\\
3200	0\\
3210	0\\
3220	0\\
3230	0\\
3240	0\\
3250	0\\
3260	0\\
3270	0\\
3280	0\\
3290	0\\
3300	0\\
3310	0\\
3320	0\\
3330	0\\
3340	0\\
3350	0\\
3360	0\\
3370	0\\
3380	0\\
3390	0\\
3400	0\\
3410	0\\
3420	0\\
3430	0\\
3440	0\\
3450	0\\
3460	0\\
3470	0\\
3480	0\\
3490	0\\
3500	0\\
3510	0\\
3520	0\\
3530	0\\
3540	0\\
3550	0\\
3560	0\\
3570	0\\
3580	0\\
3590	0\\
3600	0\\
3610	0\\
3620	0\\
3630	0\\
3640	0\\
3650	0\\
3660	0\\
3670	0\\
3680	0\\
3690	0\\
3700	0\\
3710	0\\
3720	0\\
3730	0\\
3740	0\\
3750	0\\
3760	0\\
3770	0\\
3780	0\\
3790	0\\
3800	0\\
3810	0\\
3820	0\\
3830	0\\
3840	0\\
3850	0\\
3860	0\\
3870	0\\
3880	0\\
3890	0\\
3900	0\\
3910	0\\
3920	0\\
3930	0\\
3940	0\\
3950	0\\
3960	0\\
3970	0\\
3980	0\\
3990	0\\
4000	0\\
};
\addlegendentry{ADr's P$_\text{Tx}$ = -5\,dBm}

\addplot [color=mycolor3,line width=1.0pt]
  table[row sep=crcr]{%
0	204.650268554688\\
10	211.967961472692\\
20	220.477357972899\\
30	230.468999837607\\
40	242.027880358867\\
50	255.030009850741\\
60	269.221930236867\\
70	284.316675905005\\
80	300.055913642272\\
90	316.232519670761\\
100	332.690437663323\\
110	349.315948842039\\
120	366.027198569787\\
130	382.765838616687\\
140	399.490307053147\\
150	416.170914003629\\
160	432.786697771881\\
170	449.322780396519\\
180	465.768730506483\\
190	482.117382806158\\
200	498.363721934632\\
210	514.504783437194\\
220	530.538701351821\\
230	546.464536369245\\
240	562.282091875669\\
250	577.991656458302\\
260	593.593929859928\\
270	609.089825932539\\
280	624.480529334051\\
290	639.767312143409\\
300	654.951547831219\\
310	670.0346414525\\
320	685.018111692234\\
330	699.90338770498\\
340	714.692005285062\\
350	729.385307795451\\
360	743.984894731635\\
370	758.492151762713\\
380	772.9083896291\\
390	787.235016048346\\
400	801.473289744632\\
410	815.624669402042\\
420	829.690199961451\\
430	843.671187528589\\
440	857.568747700729\\
450	871.384049895\\
460	885.11817794505\\
470	898.772151607761\\
480	912.347127411425\\
490	925.843926761323\\
500	939.263584950405\\
510	952.606968873456\\
520	965.874999992141\\
530	979.06853179135\\
540	992.1884997933\\
550	1005.23551892153\\
560	1018.21050557704\\
570	1031.11408956362\\
580	1043.94716157217\\
590	1056.71027023307\\
600	1069.40408378104\\
610	1082.02937838511\\
620	1094.58666522654\\
630	1107.07662753743\\
640	1119.49979815477\\
650	1131.85680288195\\
660	1144.14813591034\\
670	1156.37435762004\\
680	1168.53597564143\\
690	1180.63354183022\\
700	1192.66741844007\\
710	1204.63815613135\\
720	1216.54620677494\\
730	1228.39196719709\\
740	1240.17582641076\\
750	1251.89833455007\\
760	1263.55972522447\\
770	1275.16043542263\\
780	1286.70086501484\\
790	1298.18130701704\\
800	1309.60212223607\\
810	1320.96373645009\\
820	1332.26632261957\\
830	1343.5102927994\\
840	1354.69591902694\\
850	1365.82351628281\\
860	1376.89330945403\\
870	1387.90562224306\\
880	1398.86067849872\\
890	1409.75872763271\\
900	1420.60004189124\\
910	1431.38484588587\\
920	1442.11337955369\\
930	1452.78582669248\\
940	1463.40237900171\\
950	1473.9633729808\\
960	1484.46887059309\\
970	1494.91907128299\\
980	1505.31417179679\\
990	1515.65443533737\\
1000	1525.939902324\\
1010	1536.17081700477\\
1020	1546.34726597433\\
1030	1556.46946568284\\
1040	1566.5375420725\\
1050	1576.55160032342\\
1060	1586.51179751609\\
1070	1596.41819120491\\
1080	1606.27103905486\\
1090	1616.07034386141\\
1100	1625.8162293508\\
1110	1635.50886488179\\
1120	1645.14823135199\\
1130	1654.73450640594\\
1140	1664.26775329498\\
1150	1673.74807490944\\
1160	1683.17553385627\\
1170	1692.55023035211\\
1180	1701.87214069797\\
1190	1711.14143732763\\
1200	1720.35808516611\\
1210	1729.52224507289\\
1220	1738.63386577204\\
1230	1747.69300882798\\
1240	1756.69976738984\\
1250	1765.65418231705\\
1260	1774.55623993201\\
1270	1783.4060390362\\
1280	1792.20353781313\\
1290	1800.94872044326\\
1300	1809.64168290494\\
1310	1818.28246237917\\
1320	1826.87094837712\\
1330	1835.40731411715\\
1340	1843.89140993045\\
1350	1852.32328259454\\
1360	1860.70300265248\\
1370	1869.03048686535\\
1380	1877.30567239156\\
1390	1885.52869576163\\
1400	1893.69935727232\\
1410	1901.81774518197\\
1420	1909.88387873657\\
1430	1917.89761507563\\
1440	1925.85901114422\\
1450	1933.76796011291\\
1460	1941.62455635741\\
1470	1949.42854364493\\
1480	1957.18005125267\\
1490	1964.87894805787\\
1500	1972.52530539534\\
1510	1980.11883631158\\
1520	1987.65973801927\\
1530	1995.1477530798\\
1540	2002.58292145798\\
1550	2009.96520386169\\
1560	2017.29447990223\\
1570	2024.57064248123\\
1580	2031.79359775103\\
1590	2038.96336195544\\
1600	2046.07986863887\\
1610	2053.14296687926\\
1620	2060.15261535899\\
1630	2067.10858929607\\
1640	2074.01096991429\\
1650	2080.8596548774\\
1660	2087.65435494236\\
1670	2094.39518796948\\
1680	2101.08198626572\\
1690	2107.71454218507\\
1700	2114.29285857735\\
1710	2120.8167486377\\
1720	2127.2861866876\\
1730	2133.7010061572\\
1740	2140.06110072576\\
1750	2146.36632301254\\
1760	2152.6165349657\\
1770	2158.81165916118\\
1780	2164.95152510368\\
1790	2171.0360225752\\
1800	2177.06505058084\\
1810	2183.03836160183\\
1820	2188.95592376524\\
1830	2194.81755834236\\
1840	2200.62309459325\\
1850	2206.372422222\\
1860	2212.065334235\\
1870	2217.70173651772\\
1880	2223.28143767556\\
1890	2228.8043066968\\
1900	2234.27016732557\\
1910	2239.67885065991\\
1920	2245.03019514275\\
1930	2250.32399302579\\
1940	2255.56015068387\\
1950	2260.73847475361\\
1960	2265.85872477385\\
1970	2270.92077464166\\
1980	2275.92445139731\\
1990	2280.86953461086\\
2000	2285.75586454862\\
2010	2290.58323376738\\
2020	2295.35149575571\\
2030	2300.06040132869\\
2040	2304.70981684856\\
2050	2309.29945060334\\
2060	2313.82918150506\\
2070	2318.29878490326\\
2080	2322.70804205219\\
2090	2327.05679550054\\
2100	2331.34478317511\\
2110	2335.57174852445\\
2120	2339.73760759548\\
2130	2343.8420600551\\
2140	2347.88486662507\\
2150	2351.86584959769\\
2160	2355.78478118272\\
2170	2359.64143921228\\
2180	2363.43555090046\\
2190	2367.16696145584\\
2200	2370.83540924824\\
2210	2374.44063798791\\
2220	2377.98239675491\\
2230	2381.46049672319\\
2240	2384.87464126609\\
2250	2388.22465259668\\
2260	2391.51024484581\\
2270	2394.73113732811\\
2280	2397.88716877166\\
2290	2400.97806942597\\
2300	2404.00351757959\\
2310	2406.96331118654\\
2320	2409.85719646182\\
2330	2412.68486755871\\
2340	2415.44613883751\\
2350	2418.14071521624\\
2360	2420.76830685568\\
2370	2423.32862920858\\
2380	2425.82151862156\\
2390	2428.24658595755\\
2400	2430.60362071836\\
2410	2432.89236016119\\
2420	2435.11248905707\\
2430	2437.26375558829\\
2440	2439.34591355515\\
2450	2441.35860610272\\
2460	2443.30159811356\\
2470	2445.17466023747\\
2480	2446.97739404071\\
2490	2448.70958145026\\
2500	2450.37095193174\\
2510	2451.96118233257\\
2520	2453.48001366855\\
2530	2454.92713438413\\
2540	2456.30218023603\\
2550	2457.60502701594\\
2560	2458.83520531351\\
2570	2459.99254451981\\
2580	2461.07670442299\\
2590	2462.08740956581\\
2600	2463.02427334718\\
2610	2463.88709148455\\
2620	2464.67554870038\\
2630	2465.38927734796\\
2640	2466.02803374895\\
2650	2466.59152205775\\
2660	2467.07945306748\\
2670	2467.49142665092\\
2680	2467.8272258935\\
2690	2468.0865231353\\
2700	2468.26899766047\\
2710	2468.37433579299\\
2720	2468.40228984232\\
2730	2468.35244279405\\
2740	2468.22462035901\\
2750	2468.01842025836\\
2760	2467.73350659471\\
2770	2467.36966878027\\
2780	2466.92652742201\\
2790	2466.40376982085\\
2800	2465.8011500338\\
2810	2465.11831253901\\
2820	2464.354910035\\
2830	2463.51066231948\\
2840	2462.58529757517\\
2850	2461.57843504761\\
2860	2460.48982014531\\
2870	2459.31908966558\\
2880	2458.06600671629\\
2890	2456.73016752683\\
2900	2455.31135346738\\
2910	2453.80917945612\\
2920	2452.22338711445\\
2930	2450.55361067631\\
2940	2448.79961120593\\
2950	2446.96104279133\\
2960	2445.03756977695\\
2970	2443.02892520704\\
2980	2440.93473599288\\
2990	2438.75475632818\\
3000	2436.48863475299\\
3010	2434.13614718315\\
3020	2431.69684830213\\
3030	2429.17053666223\\
3040	2426.55684826127\\
3050	2423.85548901512\\
3060	2421.066176822\\
3070	2418.18858399499\\
3080	2415.22239541473\\
3090	2412.16730870216\\
3100	2409.02303438946\\
3110	2405.7892386535\\
3120	2402.46554387069\\
3130	2399.05181559817\\
3140	2395.54758916435\\
3150	2391.95264344976\\
3160	2388.26660020852\\
3170	2384.48915331249\\
3180	2380.6200685178\\
3190	2376.65901292093\\
3200	2372.60561274226\\
3210	2368.45956715805\\
3220	2364.22059156669\\
3230	2359.88841779518\\
3240	2355.4626255104\\
3250	2350.94292480287\\
3260	2346.32904326739\\
3270	2341.62061433821\\
3280	2336.81729028741\\
3290	2331.91885390847\\
3300	2326.92482841455\\
3310	2321.83497993473\\
3320	2316.64898324911\\
3330	2311.36642324852\\
3340	2305.98701688841\\
3350	2300.51039228721\\
3360	2294.93619998599\\
3370	2289.26411324236\\
3380	2283.49377373954\\
3390	2277.62479256166\\
3400	2271.65685980507\\
3410	2265.58952754399\\
3420	2259.42242836523\\
3430	2253.15527498916\\
3440	2246.78753753305\\
3450	2240.31892972869\\
3460	2233.74903234702\\
3470	2227.07740238101\\
3480	2220.30357422135\\
3490	2213.42727260649\\
3500	2206.44788284507\\
3510	2199.36514086809\\
3520	2192.17844696091\\
3530	2184.88739477797\\
3540	2177.49150877315\\
3550	2169.99029830745\\
3560	2162.38331074691\\
3570	2154.66992572155\\
3580	2146.84977025708\\
3590	2138.92220374623\\
3600	2130.88668105906\\
3610	2122.74270136993\\
3620	2114.48965688257\\
3630	2106.12693716191\\
3640	2097.65393074361\\
3650	2089.06997675828\\
3660	2080.37451704238\\
3670	2071.56684777259\\
3680	2062.64622309262\\
3690	2053.6120546809\\
3700	2044.46351865639\\
3710	2035.19985529786\\
3720	2025.82036988075\\
3730	2016.32424029829\\
3740	2006.71061854902\\
3750	1996.97863392558\\
3760	1987.1273962664\\
3770	1977.15609402323\\
3780	1967.0637124167\\
3790	1956.84932303441\\
3800	1946.51189832215\\
3810	1936.05041042943\\
3820	1925.4638348824\\
3830	1914.7510625009\\
3840	1903.91081444546\\
3850	1892.94210523405\\
3860	1881.84351684764\\
3870	1870.61393381921\\
3880	1859.25200429402\\
3890	1847.75632966535\\
3900	1836.12547172991\\
3910	1824.35804753858\\
3920	1812.45256003401\\
3930	1800.40732198499\\
3940	1788.2208128639\\
3950	1775.89125465082\\
3960	1763.41696738736\\
3970	1750.79612224581\\
3980	1738.02675593959\\
3990	1725.10695121917\\
4000	1712.03476350716\\
};
\addlegendentry{ADr's P$_\text{Tx}$ = 0\,dBm}

\end{axis}
\end{tikzpicture}%